\documentclass [11pt]{article}
\pdfoutput=1
\usepackage{graphicx}
\usepackage{amsmath,amssymb} 
\usepackage{latexsym}
\usepackage{mathrsfs}
\usepackage{color}
\definecolor{red}{rgb}{1,0,0}

\makeatletter
\def\section{\@startsection {section}{1}{\z@}{-3.5ex plus -1ex minus
 -.2ex}{2.3ex plus .2ex}{\large\bf}}
\def\subsection{\@startsection{subsection}{2}{\z@}{-3.25ex plus -1ex
minus -.2ex}{1.5ex plus .2ex}{\normalsize\bf}}
\makeatother
\makeatletter

\@addtoreset{equation}{section}

\makeatother

\def\UV{{\textsc{uv}}}
\def\IR{{\textsc{ir}}}
\def\Ls{{\textsc{l}}}
\def\Rs{{\textsc{r}}}

\def\pslash{\raisebox{1pt}{$\slash$} \hspace{-7pt} p}

\def\bea{\begin{eqnarray}} \def\eea{\end{eqnarray}}
\def\be{\begin{equation}} \def\ee{\end{equation}} \def\nn{\nonumber}
  \def\Z{{\bf Z}}

\begin{document}

\thispagestyle{empty}

\begin{center}

\begin{center}

\vspace*{0.5cm}

{\Large\bf Dark Matter and Electroweak Symmetry Breaking \\ [2mm]
in Models with Warped Extra Dimensions}

\end{center}

\vspace{1.4cm}

{\bf Giuliano Panico$^{a,b}$,
Eduardo Pont\'on$^c$, Jos\'e Santiago$^{d,e}$ and Marco
Serone$^{a}$}\\

\vspace{1.2cm}

${}^a\!\!$
{\em ISAS-SISSA and INFN, Via Beirut 2-4, I-34013 Trieste, Italy}

\vspace{.3cm}

${}^b\!\!$
{\em Physikalisches Institut der Universit\"at Bonn, Nussallee 12, 53115 Bonn, Germany}\\

\vspace{.3cm}

${}^c\!\!$
{\em Department of Physics, Columbia University,\\
538 W. 120th St, New York, NY 10027, USA}\\

\vspace{.3cm}

${}^d\!\!$
{\em FERMILAB, P.O. Box 500, Batavia, IL
  60510, USA}
\\
${}^e\!\!$
{\em Institute for Theoretical Physics, ETH, CH-8093, Z\"urich, Switzerland}\\

\end{center}

\vspace{0.8cm}

\centerline{\bf Abstract}
\vspace{2 mm}
\begin{quote}

We show that a discrete \textit{exchange} symmetry can give rise to
realistic dark matter candidates in models with warped extra
dimensions.  We show how to realize our construction in a variety of
models with warped extra dimensions and study in detail a realistic
model of Gauge-Higgs Unification/composite Higgs in which the observed
amount of dark matter is naturally reproduced.  In this model, a
realistic pattern of electroweak symmetry breaking typically occurs in
a region of parameter space in which the fit to the electroweak
precision observables improves, the Higgs is heavier than the
experimental bound and new light quark resonances are predicted.  We
also quantify the fine-tuning of such scenarios, and discuss in which
sense Gauge-Higgs Unification models result in a natural theory of
electroweak symmetry breaking.

\end{quote}

\vfill

\newpage

\section{Introduction}

Models with warped extra dimensions~\cite{Randall:1999ee} have arisen
in the last few years as strong candidates for a natural theory of
electroweak symmetry breaking (EWSB).  The original solution to the
hierarchy problem has been supplemented by the addition of a natural
flavor structure~\cite{flavor}, a custodial symmetry to protect the
$T$ parameter~\cite{Agashe:2003zs} and the $Z \bar{b}_L b_L$
coupling~\cite{Agashe:2006at}, and the realization of the Higgs as the
pseudo Goldstone Boson of a broken global
symmetry~\cite{Contino:2003ve}--\cite{Agashe:2005dk}.  These
developments have produced calculable models that successfully address
most of the mysteries related to the electroweak (EW) scale.

On the other hand, dark matter (DM), that is often considered one of
the strongest --albeit indirect-- hints of physics beyond the Standard
Model (SM), has so far lacked a generic implementation in models with
warped extra dimensions.  The main reason is the inherent asymmetry in
warped backgrounds, that do not posses a natural KK parity as the one
present in Universal Extra Dimensions (UED)~\cite{Appelquist:2000nn}.

In this article we explore a generic procedure to introduce an exact
discrete exchange symmetry that results in new stable states, without
introducing new parameters.  This is done via a doubling of part of
the field content.  The exchange symmetry we advocate has been first
introduced in \cite{Panico:2006em} (where it was dubbed ``mirror
symmetry'') and \cite{Regis:2006hc} to alleviate the fine--tuning and
to get a viable DM candidate in Gauge-Higgs Unification (GHU) models
in flat space.  As already anticipated in \cite{Panico:2006em}, it can
be extended straightforwardly to warped models.  Given a bulk field,
$\phi$, satisfying certain boundary conditions (b.c.), the procedure
consists of replacing $\phi$ by a pair of fields, $\phi_{1}$ and
$\phi_{2}$, and imposing the symmetry $\phi_1 \leftrightarrow
\phi_2$.~\footnote{A similar discrete symmetry has recently been used
in \cite{Bai:2008cf} to get a DM candidate in the context of little
Higgs theories.} The even linear combination $\phi_+ \equiv (\phi_1 +
\phi_2)/\sqrt{2}$ is identified with the original field (in
particular, it inherits the b.c. obeyed by $\phi$, as well as its
couplings).  The couplings of the orthogonal combination, $\phi_-
\equiv (\phi_1 - \phi_2)/\sqrt{2}$ are determined by those of
$\phi_{+}$.  Under the exchange symmetry one has
$\phi_{\pm}\rightarrow \pm \phi_{\pm}$, so that one can assign a
multiplicative charge $+1$ to $\phi_+$ and $-1$ to $\phi_-$.  Provided
the discrete exchange symmetry is an exact symmetry at the quantum
level, the lightest Kaluza--Klein (KK) resonance among all $\Z_2$-odd
states in the model is absolutely stable, and will be referred to as
the LOP (Lightest Odd Particle).  We argue that the above symmetry is
indeed exact, by showing that possible 5D Chern--Simons (CS) terms, in
general needed to restore gauge invariance in 5D theories, do not
violate it~\cite{Hill:2007zv}.

The choice of which fields to double must be guided by
phenomenological considerations.  For example, DM direct and indirect
searches impose stringent constraints on the possible couplings of the
DM candidate to SM fields.  It is therefore natural to look for charge
and color neutral fields that can lead to viable DM candidates.  In
fact, the models we will consider always contain $U(1)$ factors, and
it will be natural for the DM candidate to be a $U(1)$ massive gauge
field, $X_{-}$.  This is similar to the 5D UED case and the GHU model
of \cite{Regis:2006hc} in flat space, in which the DM can be identified
with the first KK mode of the hypercharge gauge field.  If the LOP
were the only $\Z_2$-odd particle, it would couple to SM fields only
via non-renormalizable interactions.  For the typical scales involved,
its annihilation rate would then be extremely small, and the resulting
thermal relic density unacceptably large.  Hence, it is necessary in
general to apply the previously described ``doubling'' construction to
additional fields, making sure that $X_-$ remains the lightest among
the $\Z_{2}$-odd particles.  We can choose b.c. for the LOP so that it
has a mass about one order of magnitude smaller than the mass of the
first $\Z_2$-even gauge resonances.  Thus, our DM candidate has
typically sub--TeV masses, which, as we will see, can range from $\sim
100$ GeV in Higgsless models up to $\sim 700$ GeV in Randall--Sundrum
(RS) models.  Together with its mass, the coupling of the LOP to SM
matter is the other crucial parameter governing its relic density.  In
all the cases we will consider, the $U(1)$ symmetry is related to the
SM $U(1)_Y$ symmetry, with a coupling constant of electroweak size,
and the $\Z_{2}$-odd fields will consist of $X_-$ and a subset of
fermion fields, typically associated with the top or bottom quarks.
It is clear that our DM candidate is a weakly interacting massive
particle (WIMP) with approximately the correct mass and couplings to
give rise to the observed DM relic density in the universe.

The construction outlined above turns out to be particularly natural
in the specific class of GHU scenarios in warped space.\footnote{It is
known that such models can also be seen, thanks to the AdS/CFT
dictionary \cite{Ads-cft}, as strongly coupled 4D composite Higgs
models.  However, since calculability requires the 5D picture, we will
mostly use the 5D language, adopting only occasionally the 4D dual
language.  } In this case, the additional set of fields introduced by
our procedure is not only minimal, but is also singled out by the role
the doubled sector plays in generating the EW scale dynamically.
Although it is not easy to use the discrete symmetry to protect EW
observables from large tree-level corrections, as in R-parity
preserving supersymmetric scenarios, Little Higgs theories with
T-parity, or UED with KK parity, we will see that the physics of EWSB
in the GHU model with the discrete symmetry naturally leads to a
region in parameter space where the constraints due to precision
measurements are relaxed.  Furthermore, this same region of parameter
space, in which coannihilation effects can be relevant, predicts a
dark matter abundance in accord with observation.  Thus, the DM sector
is tightly connected to the physics of EWSB, and plays an indirect
role in leading to agreement with precision constraints.\footnote{See
\cite{Hambye:2007vf} for a model in which EWSB and DM are related.  In
this reference, however, the hierarchy problem is not addressed.}
Also, the model is rather predictive, with several fermionic
resonances nearly degenerate with the DM particle that should lead to
an exciting collider phenomenology.

We also consider some non-perturbative effects --the formation of
bound states and the effect of QCD Coulomb-like forces-- which might
invalidate the usual perturbative computation of the relic density.
We argue that the problem of the formation of bound states does not
occur and that the effect of Coulomb--like forces on the perturbative
cross-sections is negligible.

We also discuss in some detail the degree ofÊfine-tuning involved in the EWSB pattern, and compare to supersymmetric extensions of the standard model.  It turns out that in the GHU framework (with or
without DM) the fine-tuning is somewhat worse than expected by naive
considerations.  We point out, however, that here the fine-tuning
seems to be associated with accommodating a top mass of order the EW
scale, rather than with an intrinsic tension in the Higgs sector.
When restricted to the region of parameter space with a fixed top
mass, the low-energy properties of the model turn out to be
essentially insensitive to the microscopic parameters of the model,
and therefore to the detailed properties of the new physics, which is a very
interesting feature.

Although we find the application of our construction to the GHU
framework particularly appealing, we stress that it is of more general
applicability.  We also briefly discuss the implementation of the
discrete exchange symmetry in two other models with warped extra
dimensions: the simplest RS model with the SM fermions and gauge
bosons in the bulk, and a Higgsless model.  Hence, our construction
leads to viable DM candidates in a variety of scenarios, without the
introduction of new parameters.

The organization of the paper is as follows.  In section~\ref{sect:Z2}
we review the generic properties of the exchange $\Z_2$ symmetry that
gives rise to a dark matter candidate.  In section~\ref{sec:GHU} we
discuss in detail its implementation in a model of GHU. In particular,
we study the interplay between EWSB, the EW constraints and the DM
relic abundance, the (ir)relevance of the above mentioned
non-perturbative effects, and discuss the fine-tuning.  In
section~\ref{sec:othermodels} we describe the implementation of the
discrete exchange symmetry in other models.
Section~\ref{sec:phenomenology} is devoted to particular details of DM
collider phenomenology and DM direct detection in our construction.
We comment on the issue of anomalies in section~\ref{sect:anomalies},
and conclude in section~\ref{sect:conclusions}.  We relegate some
technical details to the Appendices.

\section{Exchange $\Z_2$ Symmetry and Stable Particles\label{sect:Z2}}
\label{Z2}

The simplest way to get a stable massive particle $X_-$ beyond the SM
spectrum is obtained by introducing a discrete $\Z_2$ symmetry under
which all SM particles are even, with $X_-$ the lightest $\Z_2$-odd
particle.  For example, viable DM candidates in promising models of
new physics, such as SUSY or little Higgs models, are stable due to
$\Z_2$ symmetries (R- and T- parity~\cite{Cheng:2003ju},
respectively).  In the context of extra-dimensional theories, a
geometric $\Z_2$ symmetry called KK--parity leads to viable DM
particles in 5D \cite{Servant:2002aq} and 6D \cite{Dobrescu:2007ec}
UED scenarios.  Most extra-dimensional models which aim at stabilizing
the electroweak scale are based on warped compactifications of the
Randall--Sundrum type \cite{Randall:1999ee}, with bulk gauge and
fermion fields \cite{flavor}, where KK-parity cannot be trivially
imposed.\footnote{ Ref.~\cite{Agashe:2004ci} considered a discrete
symmetry to forbid low scale baryon number violation in Grand Unified
Models, that also gave rise to a stable DM candidate.  See also
\cite{Agashe:2007jb} for a recent attempt to include KK parity in
warped space, and \cite{DiazCruz:2007be} for some proposals of DM
candidates in composite Higgs models.} We are therefore motivated to
look for other discrete symmetries that could lead to DM candidates in
such warped scenarios.

From a model-building point of view, it is always possible to impose a
discrete symmetry {\it ad hoc} for the only purpose of getting a
stable particle, possibly with the correct properties to account for
the observed DM density.  Such constructions become far more
interesting if the DM sector is tightly connected with other sectors
of a given theory, in such a way that it leads to additional testable
predictions and/or no new parameters are introduced.  The $\Z_2$
exchange symmetry we consider belongs to this class, since no new
parameters are added.  Furthermore, in the particular class of GHU
models, which will be considered in detail in the next section, it can
also improve the EWSB pattern and lead to better agreement with
electroweak constraints.

Consider a model with warped extra dimensions and a bulk gauge
symmetry that includes a $U(1)$ factor, $\mathcal{G}\times U(1)_X$.
Our construction introduces a discrete symmetry that acts on the
$U(1)_X$ factor and leads to a stable spin-1 particle.  The group
$\mathcal{G}$ does not play any role in our construction (it will be
neutral under the discrete symmetry) and we will not discuss it any
further.  The first step is to double the $U(1)_X$ field, $U(1)_X \to
U(1)_{X_1}\times U(1)_{X_2}$.  The original $U(1)_X$ gauge boson is
identified with the symmetric combination, $X_+ \equiv
(X_1+X_2)/\sqrt{2}$, and the gauge couplings chosen as $g_{X_1} =
g_{X_2} = \sqrt{2} g_X$, so that $X_+$ couples with strength $g_{X}$.
The antisymmetric combination, $X_- \equiv (X_1-X_2)/\sqrt{2}$ is odd
under the $\Z_{2}$ exchange symmetry under which $X_{1}
\leftrightarrow X_{2}$.

As remarked in the introduction, doubling only a $U(1)$ symmetry is
not enough to obtain a realistic DM candidate.  In order to improve
this, we take a subset\footnote{Doubling the whole spectrum is
unnatural, since fermions with twisted boundary conditions can have
exponentially small masses \cite{Del Aguila:2001pu,Agashe:2004ci},
giving rise to unacceptable LOP's.  In fact, as we will see, requiring
$X_-$ to be the LOP constrains $|c|<1/2$ for the doubled fermions.} of
the fermionic fields in the model $\psi$, with $U(1)_X$ charge $Q_X$,
and double them into mirror pairs, $\psi \to \psi_{1,2}$, assigning
them $U(1)_{X_1}\times U(1)_{X_2}$ charges $(Q_X,0)$ and $(0,Q_X)$,
respectively.  Finally, undoubled fermions, $\varphi$, with charge $Q$
under the original $U(1)_X$ group, are assigned charges
$(\frac{1}{2}Q,\frac{1}{2}Q)$, which implies that they couple only to
the symmetric combination $X_+$, with charge $Q$.  Since $\psi_1$
couples only to $X_1$, while $\psi_2$ couples only to $X_2$, the
relevant 5D Lagrangian density is ${\cal L} = {\cal L}_1 +{\cal L}_2 $
with
\be
e^{-1} {\cal L}_i = -\frac 14 F_i^2 + \bar\psi_i \Big[\mathrm{i}
\gamma^M (\mathcal{D}_M+\mathrm{i} Q_X g_{X_{i}} X_{i\,M}) -
  m_i\Big] \psi_i\,, \ \ \ i=1,2\,,
\label{Lexchange}
\ee
where $F_i$ is the field strength of the gauge field $X_{i}$ with
5D coupling constant $ g_{X_{i}}$, $\mathcal{D}_M$ is the
gravitationally-covariant derivative and $e$ is the determinant of the
f\"unfbein associated to the 5D metric
\be
ds^2 = e^{-2 k y} \eta_{\mu\nu} dx^\mu dx^\nu - dy^2 = 
\left(\frac{z_\UV}{z}\right)^2 ( \eta_{\mu\nu} dx^\mu dx^\nu - dz^2)\,,
\ee
where, as usual, $\mu$ runs over the 4D directions, $0\leq y\leq L$,
$z_\UV\leq z \leq z_\IR$, $z=e^{ky}/k$.  In Eq.~(\ref{Lexchange}), we
have not explicitly written terms containing the gauge bosons of the
group $\mathcal{G}$ or the undoubled fermions.  The exchange symmetry
constraints $m_1=m_2=m \equiv c k$.  Notice that for simplicity we
omit a possible mixing term of the form $F_1F_2$ in
Eq.~(\ref{Lexchange}).  Similarly, depending on the b.c. chosen for
the $\Z_2$-odd fields, possible boundary terms can appear.  We assume
here that all these operators can be neglected, so that no new
parameters are introduced.  In terms of $\pm$ fields, we have
\bea
e^{-1}{\cal L} &= &  -\frac 14 (F_+^2+F_-^2)
+ \bar \psi_+ (\mathrm{i} \gamma^M \mathcal{D}_M - m) \psi_+
+ \bar \psi_- (\mathrm{i} \gamma^M \mathcal{D}_M - m) \psi_-
\nn \\
&
-& g_X Q_X  \gamma^M
(\bar \psi_+ X_{+\,M} \psi_+ + \bar \psi_- X_{+\,M} \psi_- +
 \bar \psi_+ X_{-\,M} \psi_- + \bar \psi_- X_{-\,M} \psi_+ )~.
\label{Aminuscouplings}
\eea
Let us now discuss the boundary conditions for the different fields.
As explained above, $X_{+}$ inherits the b.c. of the original $U(1)_X$
gauge boson (that can in general involve in a non-trivial way the
neutral gauge bosons in the group $\mathcal{G}$).  The b.c. for the
odd combination are taken to be
\begin{eqnarray}
X_{-\,\mu}= \frac{1}{\sqrt{2}}(X_{1\,\mu}-X_{2\,\mu}) &\sim& (+,-)~,
\label{Xminus}
\end{eqnarray}
where $+/-$ denote Neumann/Dirichlet b.c., respectively, with the
first/second entry in parenthesis referring to the UV/IR boundary
($X_{-\,5}$ satisfies opposite b.c.).  These b.c. allow for UV brane
localized kinetic terms characterized by a dimensionful coefficient
$r^{-}_\UV$ (in the notation of Ref.~\cite{Carena:2002dz}).  The
lightest $X_{-}$ resonance (the LOP) has mass of order
\be
m_{X_{-}} \simeq \sqrt{\frac{2}{k(L+r^{-}_\UV)}} \, \mu_\IR 
\simeq \sqrt{\frac{2}{kL}} \, \mu_\IR~, 
\label{DMmass}
\ee
where $\mu_\IR = k \, e^{-k L} = 1/z_\IR$ and the second equality
holds whenever the localized term is small.  This mass is
parametrically smaller than the KK scale $\mu_\IR$.  For example, for
values of $kL$ that solve the hierarchy problem, $m_{X_{-}}$ is about
a factor of 10 below the mass of other gauge resonances, which are of
order $m_{X_{+}} \simeq 2.5 \, \mu_\IR$.  Note that $(-,+)$ b.c. for
$X_-$ would instead give a larger mass, of order $m_{X_{+}}$.  It
would then be hard to identify $X_{-}$ as the LOP and get the correct
relic density with such a choice of b.c., without the
introduction of large brane kinetic terms.  For these reasons, we do
not consider this possibility.

Regarding the doubled fermions, $\psi_+$ satisfies the boundary
conditions of the original fermion.  The odd combination, $\psi_-$
should obey $(+,-)$ or $(-,+)$ b.c. (for one chirality, opposite for
the other one) so that no fermion zero modes are introduced in the
$\Z_2$-odd sector of the theory.\footnote{In principle, it is also
possible to introduce $\Z_2$-odd fermions with $(++)$ boundary
conditions, getting rid of the zero modes by coupling them through
mass terms to localized chiral fermions.  In the limit that these
masses become large, a description by effective b.c. as the one we are
using is appropriate.} However, the unbroken $U(1)_-$ symmetry on the
UV brane, that mixes the two mirror fermions $\psi_{\pm}$, requires
that the b.c. of $\psi_-$ be equal to that of $\psi_+$ \textit{on the
UV brane}.  The mass of the first KK mode of $\psi_-$ strongly depends
on $c$.  For $|c|\geq 1/2$, this lightest fermionic KK mode can be
lighter than $X_{-}$.  In the context of warped scenarios that solve
the flavor puzzle by fermion localization, this usually means that we
can only double the fermions associated with the top and bottom
quarks, which have $|c| < 1/2$.

These are the qualitative, model--independent features of our
mechanism to endow existing models with a stable particle that can be
a DM candidate.  In the next section, we apply this construction to a
very appealing model of EWSB in warped extra dimensions,
studying in detail the pattern of
EWSB, EW precision tests and the calculation of the DM relic density.
In section~\ref{sec:othermodels} we will briefly show how this
construction can be easily implemented in a variety of models with
warped extra dimensions.

\section{A Relevant Case: a GHU/Composite Higgs
Model\label{sec:GHU}}

Models of GHU (see e.g. \cite{Serone:2005ds} for a brief review and
further references) rely on an extended symmetry $\mathcal{G}$ broken
to a subgroup $\mathcal{H}$.  The KK gauge bosons, $A_\mu^{\hat{a}}$,
associated with the coset directions $\mathcal{G}/\mathcal{H}$ are all
massive, but the $A_5^{\hat{a}}$ towers give rise to zero-modes.  The
latter are 4D scalars and, by an appropriate choice of the symmetry
breaking pattern $\mathcal{G} \rightarrow \mathcal{H}$, can have the
correct quantum numbers to be identified as the Higgs.  The
higher-dimensional gauge symmetry leaves a remnant shift symmetry for
the Higgs~\cite{vonGersdorff:2002rg} that ensures that the Higgs
potential is finite to all orders.  This UV insensitivity guarantees
that the Higgs potential gets corrections of the order of the IR
scale, $\mu_\IR$, and not the cut-off of the theory, thus alleviating (though not solving)
the little hierarchy problem.  In this paper we will show how our
construction can be easily implemented in a fully realistic model of
GHU in warped extra dimensions.  We will consider the minimal
composite Higgs model with fermions in the fundamental representation
of $SO(5)$ (MCHM$_5$) of Ref.~\cite{Contino:2006qr}, but our
construction could be applied to other variations such as the models
considered in \cite{Carena:2007ua}.  The starting bulk gauge group is
$SO(5)\times U(1)_X$, broken to $SO(4)\times U(1)_X$ on the IR brane
(the actual symmetry on the IR brane is assumed to be $O(4)$) and to
the SM on the UV brane.  The quark sector is embedded in fundamental
representations of $SO(5)$ that decompose under $SO(4)\sim
SU(2)_L\times SU(2)_R$ as a bidoublet plus a singlet,
$\mathbf{5}=(2,2)\oplus(1,1)$.  The relevant sector for EWSB is the
third generation.  See appendix~\ref{App:Model} for a summary of the
field content of the model.

Let us now construct the simplest $\Z_2$ extension of the MCHM$_5$
model.  Following our prescription, we enlarge the gauge group by
doubling the $U(1)_X$ factor to $U(1)_{X_1}\times U(1)_{X_2}$ and
identify the original gauge boson $X$ with the even combination,
$X_+=\frac{1}{\sqrt{2}}(X_1+X_2)$.  The odd combination has $(+,-)$
boundary conditions as in Eq.~(\ref{Xminus}).  As emphasized in the
previous section, only fermions with mass parameter $|c|<1/2$ can be
safely doubled, since otherwise they could easily give rise to
charged/colored $\Z_{2}$-odd KK modes lighter than the $\Z_{2}$-odd KK
gauge bosons.  It is therefore natural to consider doubling the
multiplets associated with the top quark, since accommodating the top
mass requires $|c|<1/2$ for the associated 5D fields.  By looking at
Eq.~(\ref{fieldcomponents}), it is clear that the only field that can
be doubled without introducing new unwanted zero modes is $\xi_u$.  We
will see that this is sufficient to obtain the correct DM relic
density.  In summary, our quark sector is identical to that of the
MCHM$_5$ model, except that $\xi_u$ is replaced by two copies,
$\xi_{u_1}$, $\xi_{u_2}$, with $U(1)_{X_1} \times U(1)_{X_2}$ charges
$(2/3,0)$ and $(0,2/3)$, respectively, but otherwise identical in
boundary conditions and bulk mass.  The physical combinations are the
$\Z_{2}$-even $\xi^+_{u}$ (which, for simplicity, we will call
$\xi_{u}$ in what follows) and the $\Z_{2}$-odd $\xi^-_{u}$.  The
other quark and lepton fields are assigned the same charge under both
abelian groups, $(\frac{1}{2}Q_X,\frac{1}{2}Q_X)$, where $Q_X=2/3$ for
$\xi_{q_1}$, $Q_X=-1/3$ for $\xi_{q_2}$, $\xi_{d}$, and analogous
assignments for the lepton sector.  Therefore, the spectrum in our
model contains a set of fields that are even under the $\Z_2$ symmetry
and corresponds exactly to the spectrum in the original MCHM$_5$
model, plus two 5D $\Z_2$-odd multiplets, $X_-$ and $\xi^-_{u}$.  The
$\Z_{2}$-odd gauge boson has a first KK mode, which we will simply
call $X_{-}$, with a mass given by Eq.~(\ref{DMmass}).  The
$\Z_{2}$-odd fermions have first KK modes with $c_u$-dependent masses
that are, neglecting EWSB effects and provided $|c_u|<1/2$, always
larger than the one of $X_-$.\footnote{Note that odd-fermion masses
are independent of the localized mixing masses, and are therefore
entirely determined by $c_u$, up to EWSB effects.  After the mixing
due to EWSB, these could become lower than $m_{X_{-}}$ if $|c_u|$ is
very close to 1/2.} As we take $c_u \to 1/2$ the first KK mode of
$(2,2)^{u_-}$ becomes degenerate with $X_{-}$, whereas in the $c_u \to
-1/2$ limit it is $(1,1)^{u_-}$ that becomes degenerate with $X_{-}$
(see appendix~\ref{App:Model} for the notation).

\subsection{Electroweak Symmetry Breaking and Precision Constraints}
\label{sec:EW}

The pattern of EWSB in the MCHM$_5$ model and its main constraints
have been studied
in~\cite{Contino:2006qr,Carena:2007ua,Medina:2007hz}.  A realistic
pattern of EWSB can be obtained for $|c_{q_1}| \lesssim 0.4$ and $0.35
\lesssim |c_u| \lesssim 0.45$.  Outside these two regions, it is
difficult to obtain a reasonable value of the gauge boson, top and
Higgs masses.  Of particular interest is the dependence on $c_{u}$.
For fixed values of the other parameters, smaller values of $|c_u|$
result in no EWSB, while larger values of $|c_u|$ give the wrong EWSB
pattern.  The qualitative features of this dependence can be easily
understood in the limit that the localized mixing masses vanish.  In
this simplified case, the two chiralities of the top quark arise from
$\xi_{q_{1}}$, and obtaining a top Yukawa coupling of order one fixes
$c_{q_{1}} \approx 0.44$.  The contribution to the Higgs potential due
to the $\xi_{q_{1}}$ KK tower destabilizes the origin, while the
contributions from the gauge KK towers tend to align the vacuum along
the EW symmetry preserving direction.  It turns out that the top tower
contribution is so large that it would drive the vacuum expectation
value (VEV) to its maximum value, $s_{h} = 1$, where
\be
s_{h} = \sin\left( {\frac{\langle h \rangle}{f_{h}}} \right)~,
\hspace{1cm}
f_{h} = \frac{1}{g} \sqrt{ \frac{2}{kL}} \, \mu_\IR~.
\label{sh}
\ee
Here $f_{h}$ is the ``Higgs decay constant'' and $g$ is the SM
$SU(2)_{L}$ gauge coupling.  The resulting EW symmetry breaking
pattern is unacceptable since it leads to vanishing fermion masses,
and highly non-linear couplings of the Higgs to the gauge bosons that
are ruled out by EW precision measurements.  
However, fermions with ``twisted'' boundary conditions, such as $\xi_{u}$,
give a contribution to the Higgs potential that tends to align the
vacuum along the EW symmetry preserving direction.
This
contribution is controlled only by $c_{u}$ in the above simplified
limit, and turns off when $|c_{u}| \gtrsim 1/2$.  If $|c_{u}| \ll
1/2$, then this EW restoring contribution overwhelms the contribution
due to $\xi_{q_{1}}$ and results in $s_{h} = 0$.  The upshot is that
the desired EWSB VEV, $0< s_{h} < 1$, is obtained for $|c_{u}| \sim
1/2$, but not necessarily too close to $1/2$.  This also illustrates
the crucial role that $\xi_{u}$ plays for EWSB. When the mixing masses
are turned on the details are more complicated, but the qualitative
features remain the same.  See appendix~\ref{HiggPotential} for
details on the computation of the Higgs potential.

On the other hand, EW precision data typically prefer values of
$|c_u|$ close to 1/2.  This is due to a sizable and negative one-loop
contribution to the Peskin-Takeuchi \cite{Peskin:1991sw} $T$ parameter
in most regions of parameter space \cite{Carena:2006bn}, together with
a non-negligible tree-level \textit{positive} contribution to the $S$
parameter.  The tree-level contribution to $S$, when the light
fermions are localized close to the UV brane, is given by
\cite{Agashe:2004rs}
\be
S_{\rm tree} \approx 
\frac{6\pi s^{2}_{h}}{g^{2} kL} \approx 
\frac{3\pi v^{2}}{\mu^{2}_\IR}~,
\ee
where the second equality holds whenever $v = \langle h \rangle \sim
174~{\rm{GeV}} \ll f_{h}$.  The negative contribution to the $T$
parameter arises at one-loop order from the lightest charge-$2/3$
members of the $SU(2)_{L} \times SU(2)_{R}$ bidoublets, and is
dominant away from the $|c_u| \sim1/2$ region.  Additionally, the
pseudo-Goldstone nature of the Higgs leads to a further negative
(positive) contribution to the $T$ ($S$) parameter, due to the
anomalous gauge couplings of the Higgs~\cite{Sakamura:2006rf}.  This
can be described by an effective Higgs mass \cite{Barbieri:2007bh}
\be
m_{h,\textrm{eff}} = m_{h} \left( \frac{\Lambda}{m_{h}} \right)^{s^2_{h}}~,
\label{mheff}
\ee
where $\Lambda$ is an effective cutoff scale of the order of the mass
of the first SM gauge KK resonances, $m_{KK}$, so that the
corresponding shifts in $S$ and $T$ are
\begin{eqnarray}
\Delta S_h &=& \frac{1}{12 \pi} 
\ln \left(\frac{m^{2}_{h,\textrm{eff}}}{m_{h_{\rm ref}}^2}\right)
\, ,
\nonumber\\
\Delta T_h &=& -\frac{3}{16 \pi c^2_{W}} 
\ln \left(\frac{m^{2}_{h,\textrm{eff}}}{m_{h_{\rm ref}}^2}\right)
\, .
\nonumber
\end{eqnarray}
Here $m_{h_{\rm ref}}$ is the reference Higgs mass used in the
$S$-$T$ fit to the EW data, and $c_{W}$ is the cosine of the
Weinberg angle.

The $S$-$T$ analysis reveals that in the $\Z_{2}$ extended MCHM$_{5}$
model, the regions with $|c_{u}| \sim 1/2$ are preferred (see
Fig.~\ref{fig:constraintscupos}).  For $c_{u} \sim -1/2$, the reason
can be traced back to the fact that the first KK excitation of the
$SO(4)$ \textit{singlet} in $\xi_{u}$, that mixes with the top quark,
becomes light and gives a positive contribution to the $T$ parameter
that compensates the negative contributions coming from the lightest
bidoublet states and effective heavy Higgs, Eq.~(\ref{mheff}).  As a
result, one finds a sizable region in parameter space compatible with
EWSB, where the EW precision measurements are relaxed.  For $c_{u}
\sim 1/2$, it is again possible to satisfy the EW constraints,
although the reason is somewhat more involved than for $c_{u} \sim
-1/2$.  As emphasized in \cite{Carena:2006bn}, the charge-$2/3$
members of the bidoublets give rise to both positive (from the
$T^3_{L} = +1/2$, $T^3_{R} = -1/2$ states) and negative (from the
$T^3_{L} = -1/2$, $T^3_{R} = +1/2$ states) contributions to $T$.  In
the custodially symmetric limit these two contributions would cancel
exactly, but in most regions the fact that the $T^3_{L} = -1/2$ state
is lighter than the $T^3_{L} = +1/2$ one, typically results in a net
negative contribution to $T$ (the signs are simply determined by the
quantum numbers).  However, when $c_{u} \sim 1/2$, the lightest
bidoublets arise mostly from $\xi_{u}$ (and therefore have a nearly
flat wavefunction component, that vanishes on the IR brane), and lead
to a contribution to $T$ via mass mixing with the bidoublet in
$\xi_{q_{1}}$.  It is then possible to suppress the coupling to the
Higgs of the lighter $T^3_{L} = -1/2$ state by ensuring that it lives
mostly in the component with the nearly flat wavefunction, while the
heavier $T^3_{L} = +1/2$ state has a larger component in the bidoublet
of the $\xi_{q_1}$ multiplet, that is localized near the IR brane.  In
this way, the positive contribution due to the $T^3_{L} = +1/2$ state
can dominate, and explains how the $c_{u} \sim 1/2$ region can be
compatible with the EW precision data with a relatively low
scale.\footnote{This important positive contribution is not present in
the minimal models studied in Ref.~\cite{Barbieri:2007bh} (it is,
however, in models with a custodial protection of the $Z \bar{b}_L b_L
$ coupling as the one we are considering).}

In addition to the oblique corrections parametrized by $S$ and $T$,
non-oblique corrections associated with the third generation can also
be relevant.  These can be separated into flavor-preserving versus
flavor-violating effects.  A complete treatment of the latter would
require the specification of the flavor structure of the model, which
is beyond the scope of this paper.  Most notable among the
flavor-preserving effects are the corrections to the
$Z\bar{b}_{L}b_{L}$ vertex, which has been measured at the few per
mille level.  Here we notice that the present model enjoys the
custodial protection pointed out in \cite{Agashe:2006at}, which
reduces the tree-level corrections to this vertex to a level well
below the experimental precision.  However, as first pointed out in
\cite{Carena:2006bn,Carena:2007ua} (see also \cite{Barbieri:2007bh})
the one-loop contributions to this vertex can be significant and
correlated with the one-loop contribution to the $T$ parameter
discussed above.  Hence, we also include these effects in the fit to
the EW precision observables, although, for the reasons explained
above, we do not include flavor violating
effects.\footnote{Nevertheless, we have checked that a subset of the
flavor-violating corrections, that are closely connected to parameters
entering in the flavor-preserving effects~\cite{Barbieri:2007bh}, do
not significantly alter the bounds.  Specifically, these include
certain loop-level contributions to the $Z \bar{b} s$ vertex that are
bound by $B \rightarrow X_{s} l^+l^-$ decays (these are likely to be
more important in the $c_{u} < 0$ region), as well as tree-level
contributions to the $W \bar{t}_{R} b_{R}$ vertex, that can be bound
from the $B \rightarrow X_{s} \gamma$ branching ratio
\cite{Lari:2008iz} (these affect mostly the $c_{u} > 0$ region).} In
this analysis we have performed a fit to the $Z$-pole observables and
the $W$ mass \cite{Han:2004az}, and include the effects of four-fermion interactions
that enter through the Fermi constant (these effects are subdominant).
The upshot is that there is a bound of about $\mu_\IR\sim 1.3~{\rm
TeV}$ in the $c_{u} > 0$ region, while the $c_{u} < 0$ region is
somewhat more constrained, with a lower bound $\mu_\IR\sim 1.7~{\rm
TeV}$.

Notice that although the $\Z_{2}$-odd fermions do not contribute to
$T$ (since the b.c. for $\xi^-_{u}$ respect the custodial symmetry
exactly and there is no mixing with custodial violating sectors of the
theory due to the exact $\Z_{2}$ symmetry), they affect the
minimization of the Higgs potential in an analogous way to $\xi_{u}$,
discussed at the beginning of this section.  In particular, the
existence of these additional fields satisfying ``twisted'' b.c., and
whose effects on the Higgs potential are controlled also by $c_{u}$,
means that the minimization of the potential favors values of
$|c_{u}|$ that are closer to $1/2$ than if the $\Z_{2}$-odd fermions
were absent.  This is a welcome feature, since it goes in the
direction preferred by the EW precision data, as discussed above.
Furthermore, the opening of a well defined region with $c_{u} \sim
1/2$, where the $(t_{L},b_{L})$ doublet is more fundamental
[see $f_L^+(z)$ in Eq.~(\ref{xiuwavefunctions})] is also welcome
since it might be argued, based on flavor considerations, that such a
situation is more natural 

Thus, we find that the regions that lead to a correct EWSB pattern and
good agreement with the EW precision data largely overlap in the
$\Z_{2}$ extension of the MCHM$_{5}$ model.  We postpone a discussion
of the details associated with Fig.~\ref{fig:constraintscupos} to the
next section, after we have discussed the computation of the DM relic
density in this scenario.  For the moment, let us mention that we
predict relatively light vector-like quarks (from both $\xi_{u}$ and
$\xi^{-}_{u}$) with masses close to $m_{X_{-}} \sim \sqrt{2/k L}\,
\mu_\IR$, a very distinctive signature of these models.  It should
also be emphasized that our model has the same number of parameters as
the MCHM$_5$ without DM.

\subsection{Calculation of the Relic Abundance}
\label{sec:DMdensity}

As was discussed in the previous section, a realistic pattern of EWSB
requires $|c_u| \lesssim 1/2$.  Thus, we have two separate regions
that, \textit{a priori}, could give rise to phenomenologically relevant
scenarios.  This can have important consequences for the computation
of the DM relic abundance since, as remarked above, some of the
$\Z_{2}$-odd fermions become degenerate with the DM candidate when
$|c_{u}| = 1/2$.  In the case $c_u \sim -1/2$, it is the first KK mode
of the $SO(4)$ singlet component of $\xi^-_{u}$ that becomes close in
mass to $X_{-}$ (the LOP), while the KK modes of the bidoublet are at
least a factor of ten or so heavier.  In the other relevant region of
parameter space, $c_u \sim 1/2$ (composite $t_R$), the situation is
inverted, with the first KK modes of the members of the bidoublet of
$\xi^-_{u}$ becoming light, while the singlet is considerably heavier.
These $\Z_{2}$-odd quarks couple to $X_{-}$ and SM fermions (top or
bottom), and we will refer to the lightest of them as the NLOP
(Next-to-Lightest Odd Particle).  Depending on the degree of
degeneracy, coannihilation effects can be relevant to obtain the DM
relic abundance.  Note that for $c_{u} > 0$, there are several states
whose masses are split only by EWSB effects, and all of them can
affect the final DM relic density.

In order to get an order of magnitude estimate, we start by assuming
that coannihilation processes can be neglected (a reasonable
approximation whenever the $\Z_2$-odd quarks are heavier than $X_{-}$
by a factor of $\gtrsim 15\%$).  In this case, the only relevant
process for the calculation of the DM relic abundance is the
annihilation of $X_{-}$ pairs into SM quarks, via NLOP exchange.
Up to EWSB effects, which can be shown to give negligible
corrections, the relevant 5D interactions are
\begin{equation}\label{gRcoupling}
\overline{(1,1)}^{u}_{R}
\hspace{-2pt}\not \hspace{-2.7pt} X_-
(1,1)^{u_-}_R +\mathrm{h.c.},
\end{equation}
for $c_u \sim -1/2$, and
\begin{equation}\label{gLcoupling}
\overline{(2,2)}^{u}_{L}
\hspace{-2pt}\not \hspace{-2.7pt} X_-
(2,2)^{u_-}_L +\mathrm{h.c.},
\end{equation}
for $c_u \sim 1/2$.  Here $(1,1)^{u}_{R}$ and $(2,2)^{u}_{L}$ contain,
respectively, the right-handed (RH) top and the left-handed (LH) top and bottom zero modes.  We
denote by $\tilde g_R$ and $\tilde g_L$ the 4D couplings between the
LOP, NLOP and top/bottom quarks, arising from Eqs.~(\ref{gRcoupling})
and (\ref{gLcoupling}), respectively.  These are the crucial couplings
entering in the self-annihilation cross section of $X_-$.  In the
absence of UV brane kinetic terms, both $\tilde g_L$ and $\tilde g_R$
are bound from above as follows:
\begin{equation}
|\tilde{g}_{L,R}| \lesssim \frac{2}{3} g_X = \frac{2}{3} \frac{g
  g^\prime}{\sqrt{g^2-g^{\prime\, 2}}} \approx 0.28~,
  \label{gxbound}
\end{equation}
where $g_X$ is the $U(1)_X$ gauge coupling that we have also written
in terms of the SM $SU(2)_L\times U(1)_Y$ gauge couplings.  We refer
the reader to appendix~\ref{coupling} for a detailed derivation of the
couplings $\tilde g_L$ and $\tilde g_R$.  They are typically of the
same size, so that the main difference between the above two regions
is the presence of one decay channel (into $t_R$) for $c_u<0$, and two
(into $t_L$ and $b_L$) in the $c_u>0$ case.  For this reason, it will
be useful in the following to define the parameter $\eta$, such that
$\eta = 1$ for $c_u<0$ and $\eta=2$ for $c_u>0$, and write the
coupling simply as $\tilde g$, with the understanding that $\tilde g =
\tilde g_R$ for $c_u <0$ and $\tilde g = \tilde g_L$ for $c_u >0$.

The computation of the DM relic abundance is standard (see
e.g.~\cite{Kolb:1990vq}).  In the freeze-out approximation, it can be
written in terms of the coefficients of the non-relativistic expansion
of the 
annihilation cross section, $\sigma(X_{-}
X_{-} \to SM)$,
\begin{equation}
v \sigma = a + v^2 b + \ldots,
\end{equation}
as
\begin{equation}
\Omega h^2 \approx \frac{1.04 \times 10^9}{M_P}
\frac{x_F}{\sqrt{g_\ast}} \frac{1}{a+3 b/x_F}~,
\end{equation}
where $M_P \approx 1.22 \times 10^{19}$ GeV is the Planck mass,
$x_F=m_{X_-}/T_F$, with $T_F$ the freeze-out temperature (in our case,
with $\lesssim $ TeV weakly interacting particles, $x_F\sim 24-26$),
$g_\ast$ is the effective number of relativistic degrees of freedom at
freeze-out ($g_\ast = 86.25$ for $100$ GeV $\lesssim m_{X_{-}}
\lesssim $ TeV) and $a,b$ are measured in ${\rm GeV}^{-2}$.
Neglecting EWSB effects, the annihilation cross section depends on
$\tilde{g}$, $m_{X_{-}}$ and the mass of the lightest modes in
$\xi^-_{u}$, which we call $m_{\psi}$.  It turns out that the factor
$b$ gives a negligible effect.  Denoting
\be
\Delta\equiv \frac{m_{\psi}-m_{X_-}}{m_{X_-}}~,
\label{Delta}
\ee
we obtain
\begin{eqnarray}
a&=&
\frac{2 \,\eta \, \tilde{g}^4 }{3\pi} \frac{1}{(1+(1+\Delta)^2)^2}\,
m_{X_{-}}^{-2}~,
\end{eqnarray}
and
\begin{equation}
\Omega h^2 \approx \frac{0.15}{\eta}  \left(\frac{m_{X_{-}}}{400 \,
    \mathrm{GeV}}\right)^2
\left(\frac{0.28}{\tilde{g}}\right)^4
\left(\frac{1+(1+\Delta)^2}{1+(1+0.15)^2}\right)^2.
\end{equation}
This result is actually very accurate.  Using the full expression for
the annihilation cross section and the freeze-out temperature, we
obtain, again assuming that the difference in mass between $X_{-}$ and
the $\Z_2$-odd quarks is $\gtrsim 15\%$,
\begin{equation}
\Omega h^2 \gtrsim
\left\{
\begin{array}{ll}
0.16 ~(0.6),& \quad c_u <0~, \\
0.08 ~(0.28),& \quad c_u>0~,
\end{array}
\right .
\label{DMnocoannihilations}
\end{equation}
where we assumed $m_{X_{-}} \gtrsim 400$ GeV, and the numbers are for
$\tilde{g}=0.28$ (maximal coupling) and, in parenthesis, for
$\tilde{g}=0.2$.  We therefore see that the observed relic
abundance~\cite{Komatsu:2008hk},
\begin{equation}
\Omega_{DM} h^2=0.1143 \pm 0.0034,
\label{WMAP}
\end{equation}
can be accounted for with just $X_{-}$ annihilation for $c_u>0$,
provided $\tilde g$ is large enough.  For $c_u<0$, the observed DM
energy density seems more difficult to accommodate in the simplest
scenario where only $X_{-}$ annihilations are relevant.\footnote{We
note that the presence of kinetic mixing between the two $U(1)$
factors, or UV brane localized kinetic terms for them, can change the
coupling relevant for annihilation of LOP's, leading to a realistic DM
relic abundance.  We assume that such terms are small, so that no
relevant additional parameters are introduced.} However, as explained
above, EWSB and the EW precision measurements independently point to a
region in parameter space with one or more vector-like quarks that are
nearly degenerate with the LOP, and a proper determination of the
relic abundance should take coannihilation effects into
account~\cite{Griest:1990kh}.  This is welcome because, due to the
colored nature of the new particles, coannihilations tend to increase
the cross section and therefore decrease the relic abundance to values
compatible with observation for smaller couplings and/or larger
$m_{X_{-}}$.

The relevant processes that enter the relic abundance computation are
$X_{-} X_{-} \rightarrow t \bar{t} \, (b \bar{b})$, $X_{-} \psi_{-}
\rightarrow g t \, (g b)$, $X_{-} \bar{\psi}_{-} \rightarrow g \bar{t}
\, (g \bar{b})$, $\psi_{-} \bar{\psi}_{-} \rightarrow q \bar{q}$ and
$\psi_{-} \bar{\psi}_{-} \rightarrow g g$, where $\psi_{-}$ stands for
any of the $\Z_{2}$-odd fermions that are nearly degenerate with
$X_{-}$, $g$ is the SM gluon and $q \bar q$ are SM quark-antiquark
pairs.  The relevant effective annihilation cross section reads
\begin{eqnarray}
\sigma_{\rm eff} &=& \frac{1}{g_{\rm eff}^{2}} 
\left[ \rule{0mm}{7mm}
9 \, \sigma_{X_{-} X_{-} \rightarrow t \bar{t}} 
+9(\eta-1) \, \sigma_{X_{-} X_{-} \rightarrow b \bar{b}} 
\right. \nonumber \\
&&\hspace{7mm} \mbox{} 
+72 \, e^{-x\Delta} (1+\Delta)^{3/2} \sigma_{X_{-} \psi_{-} \rightarrow g t} 
+72 (\eta-1)
\, e^{-x\Delta} (1+\Delta)^{3/2} \sigma_{X_{-} \psi_{-} \rightarrow g b} 
 \nonumber \\
& & \left. \hspace{7mm} \mbox{} 
+72 \,\eta^2\,  e^{-2x\Delta} (1+\Delta)^{3} \left( \sum_{q} \sigma_{\psi_{-} \bar{\psi}_{-} 
\rightarrow q \bar{q}} + \sigma_{\psi_{-} \bar{\psi}_{-} \rightarrow g g} \right)
\right]~,
\label{crosssectionSingletBidoublet}
\\
g_{\rm eff} &=& 3 + 12 \,\eta^2 \, e^{-x \Delta} (1+\Delta)^{3/2}~,
\nonumber 
\end{eqnarray}
where $x = m_{X_{-}}/T$, and $\Delta$ was defined in
Eq.~(\ref{Delta}).  In Eq.~(\ref{crosssectionSingletBidoublet}), the
factor $\eta$ introduced before takes into account that in the $c_{u}
< 0$ region there is a single $\Z_{2}$-odd fermion with $m_{\psi_{-}}
\approx m_{X_{-}}$, whereas for $c_{u} > 0$ there are four fermions
with $m_{\psi_{-}} \approx m_{X_{-}}$.  When $c_u>0$,
Eq.~(\ref{crosssectionSingletBidoublet}) is valid in the limit in
which we neglect the mass splitting between the fermions of the
bidoublet (which arises only from EWSB effects) and it is understood
that the cross sections with final state top (bottom) quarks involve
the $Q=2/3$ ($Q=-1/3$) heavy vector-like fermions.

\begin{figure}[t]
\centerline{ \hspace*{-0.5cm}
\includegraphics[width=0.487 \textwidth]{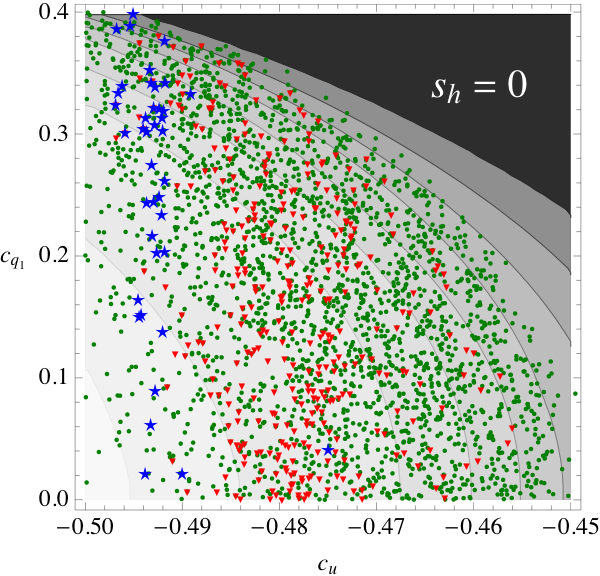}
\hspace*{0.5cm}
\includegraphics[width=0.48 \textwidth]{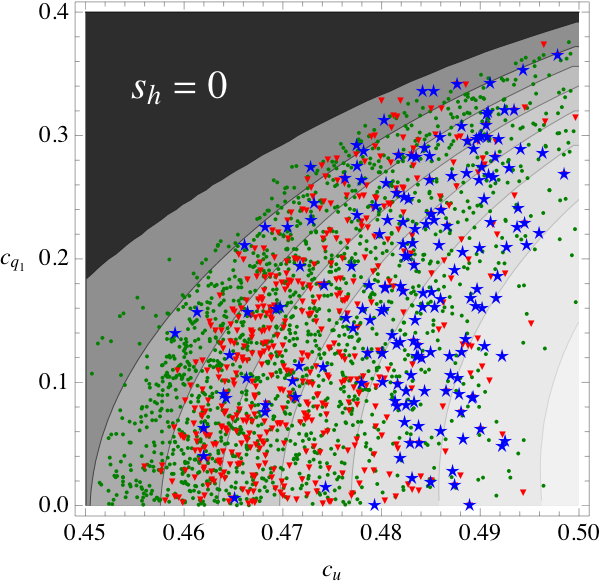}
} 
\caption{Projection onto the $c_{u}$-$c_{q_{1}}$ plane for the two
phenomenologically viable regions.  The darker areas marked $s_{h} =
0$ correspond to no EWSB. We plot $s_{h}$ averaged over the rest of
the parameters, which increases as the gray bands become lighter (see
text).  The green dots satisfy $145~{\rm GeV} < m_{t}(\mu \sim
\mu_{IR}) < 155~{\rm GeV}$.  The red triangles mark the points
consistent with the WMAP constraint, Eq.~(\ref{WMAP}), at the $2\sigma$
level.  The blue stars correspond to a sample of points that are
consistent with EW precision data at the $99\%$ CL, and the Higgs LEP bound.}
\label{fig:constraintscupos}
\end{figure}

We performed exhaustive scans, computing the pattern of EWSB, the fit
to EW precision data, the DM relic abundance and the spectrum of light
quarks.  In order to more easily automatize the scan over large
regions of parameter space while taking into account coannihilations,
we found it useful to implement the relevant features of our model in
micrOmegas~\cite{Belanger:2006is}.  This also allows us to easily take
into account the mass splittings arising from EWSB among the lightest
bidoublet states that are relevant in the $c_{u} \sim 1/2$ region, as
well as non-zero final state masses (for the typical scales of
$m_{X_{-}} \sim 400~{\rm GeV}$, the top mass can be important).  We
have checked independently that we reproduce the micrOmegas results
for the relic density to within 10\% using
Eq.~(\ref{crosssectionSingletBidoublet}), in the limit that EWSB mass
splitting effects are neglected, and in the freeze-out approximation.

We present the results separately for the two phenomenologically
interesting regions discussed previously, namely $c_{u} \sim -1/2$ and
$c_{u} \sim 1/2$.  We scanned over the following region in parameter
space: $c_{q_{1}} \in [0,0.4]$, $c_{q_{2}} \in [0.4,0.5]$, $|c_{u}|
\in [0.45,0.5]$, $c_{d} \in [-0.52,-0.45]$, $|m_{u}| \leq 1$, $|M_{u}|
\leq 1$, $|m_{d}| \leq 0.5$, $|M_{d}| \leq 0.5$, with $kL \approx 34$
in order to explain the Planck-weak scale hierarchy.  We show the
results of the scan as a projection onto the $c_{u}$-$c_{q_{1}}$ plane
in Fig.~\ref{fig:constraintscupos}, where the red triangles correspond
to the data consistent with the three-year WMAP results,
Eq.~(\ref{WMAP}), at the $2\sigma$ level.

We also highlight in the plot the data that have $145~{\rm GeV} <
m_{t}(\mu \sim \mu_\IR) < 155~{\rm GeV}$ (green dots).  As we will
discuss further in section~\ref{sec:finetuning}, the top mass plays a
rather important role in leading to a vacuum with appropriate
characteristics in these scenarios.  In the background of the figure
we exhibit in gray tones information about the degree of EWSB as
measured by $s_{h}$ [see Eq.~(\ref{sh})].  In order to make the
projection onto the $c_{u}$-$c_{q_{1}}$ plane we show the average of
$s_{h}$, computed over the rest of the parameters in the scan.  In
particular, in the darker area marked as $s_{h} = 0$, all points in
the scan lead to an EW symmetry preserving vacuum.  For $c_{u} < 0$
($c_{u} > 0$), the degree of EWSB, in the above sense, increases as
one moves towards the bottom left (right) corner of the figure
(lighter grays correspond to larger average $s_{h}$).  We therefore
see that in most of the region with non-trivial EWSB the correct top
mass can be reproduced.  Even more interestingly, the area selected by
the observed DM abundance falls in precisely the same region.  This is
related to the fact that coannhilations play a relevant role in
lowering the DM abundance to the observed level compared to the
estimates Eq.~(\ref{DMnocoannihilations}) and, as explained before,
this happens naturally in the above region (we did not compute the DM
relic abundance for points with $s_{h} = 0$).

\begin{figure}[t]
\centerline{ \hspace*{-0.5cm}
\includegraphics[width=0.47 \textwidth]{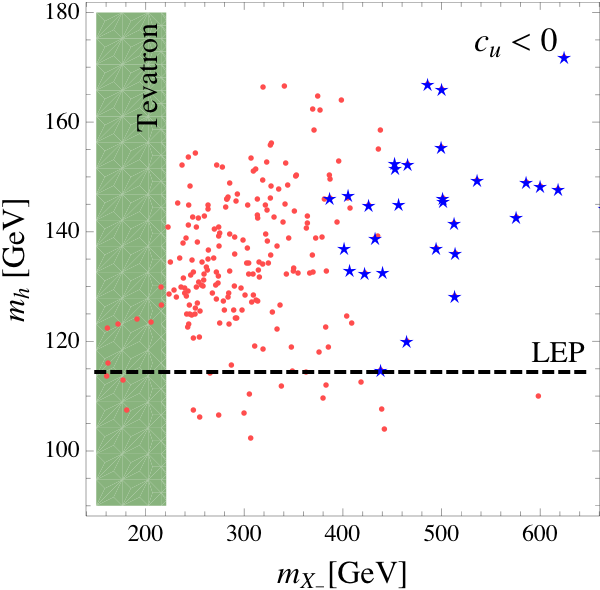}
\hspace*{0.5cm}
\includegraphics[width=0.47 \textwidth]{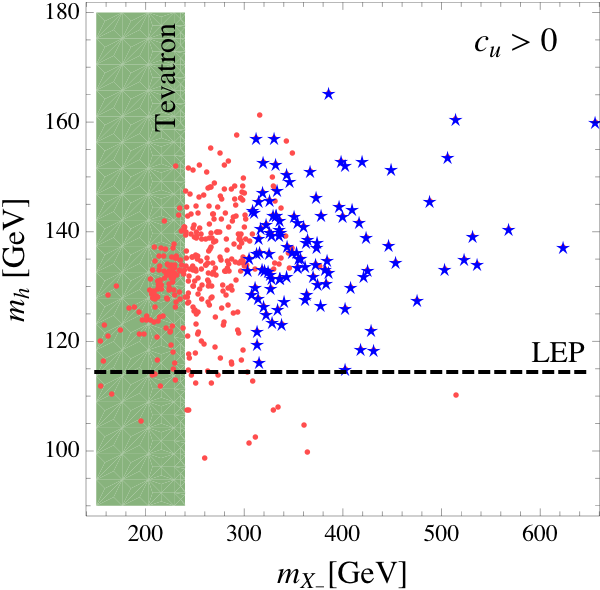}
} \caption{Masses of the DM candidate, $X_{-}$, and the Higgs for
points (red dots) that reproduce the top mass and the WMAP constraint,
Eq.~(\ref{WMAP}), for the $c_{u} < 0$ region (left panel) and $c_{u} >
0$ region (right panel).  The blue stars correspond to the subset of
points that also obey the EW precision constraints at the $99\%$ CL,
and the LEP bound on the Higgs mass, which is also indicated by the
dashed horizontal line.  The light bands indicate the approximate
bounds from the Tevatron on the colored vector-like quarks, that have
mass close to $m_{X_{-}}$, taking into account the different
multiplicities for positive and negative $c_u$.}
\label{fig:mhmD}
\end{figure}

Finally, we also show in the figure information about the fit to the
EW precision data.  As explained in section 3.1, we include in our fit
the universal corrections described by the $S$ and $T$ parameters and
the correction to the $Z\bar{b}_{L}b_{L}$ vertex (both tree- and
one-loop contributions), performing a $\Delta \chi^2$ cut at 99\% CL
with 3 d.o.f. Since this analysis is computationally intensive, we
performed the EW test only for points that satisfy $149.5~{\rm GeV} <
m_{t}(\mu \sim \mu_{IR}) < 150.5~{\rm GeV}$ (without imposing the WMAP
constraint), and for $145~{\rm GeV} < m_{t}(\mu \sim \mu_{IR}) <
155~{\rm GeV}$ when the WMAP constraint, Eq.~(\ref{WMAP}), is imposed.
The selection of the very narrow range in the first case is done only
in order to reduce the number of points to be analyzed, but it should
nevertheless give a clear picture of the situation.  We also require
that the Higgs mass be above the LEP bound, $m_{h} > 114.4~{\rm GeV}$.
Requiring that the DM relic abundance be in the range (\ref{WMAP}), we
find that in the $c_{u} > 0$ ($c_{u} < 0$) region about $25\%$
($15\%$) of the points pass the EW precision test at the $99\%$ CL.
Thus, we conclude that in the present model, the correct EWSB pattern,
consistency with the EW precision measurements, and the correct DM
abundance all occur in a common region of parameter space.  This leads
to a rather compelling picture with a light neutral spin-1 DM
candidate, and one or more fermionic resonances that are nearly
degenerate with the DM particle.  We will briefly discuss their
collider phenomenology in section~\ref{sec:phenomenology}.

Two important physical observables that are relatively well predicted
in our scenario are the Higgs mass and the mass of the DM candidate.
We show in Fig.~\ref{fig:mhmD} the distribution of points with a
realistic EWSB pattern (including the top mass) that reproduce the
observed DM relic density, in the $m_{X_{-}}-m_{h}$ plane.  We also
highlight the subset of points that are consistent with the EW
precision constraints at the $99\%$ CL. We see that the LEP bound on
the Higgs mass is easily evaded in our scenario.  Also, the Higgs
boson is expected to be lighter than about $170~{\rm GeV}$.  The mass
of the DM particle is expected to be somewhat above $300~{\rm GeV}$
($400~{\rm GeV}$) for $c_{u} > 0$ ($c_{u} < 0$).  As was mentioned
above, there is a number of vector-like quarks nearly degenerate with
the LOP. We also indicate in Fig.~\ref{fig:mhmD} the approximate
direct Tevatron bound on such particles.  Although the direct bound
already excludes a significant region in parameter space, the EW
precision analysis still puts stronger constraints on the masses of
these particles.  It is also useful to remember that, for $kL \approx
34$, we have $m_{X_{-}} \approx 0.24 \, \mu_{IR}$, which shows that
$\mu_{IR} \gtrsim 1.3~{\rm TeV}$ ($\mu_{IR} \gtrsim 1.7~{\rm TeV}$)
for $c_{u} > 0$ ($c_{u} < 0$).  Recall also that, in the absence of
brane kinetic terms, other spin-1 resonances have a mass $m_{KK}
\approx 2.5 \, \mu_{IR} \gtrsim 3.1~{\rm TeV}$ ($m_{KK} \gtrsim
4.2~{\rm TeV}$) for $c_{u} > 0$ ($c_{u} < 0$).

\subsection{Non-perturbative Corrections}

There are various physical processes that might possibly invalidate
the standard computation of the relic density based on perturbative
averaged cross-sections, namely the formation of bound states and
higher order corrections relevant for non-relativistic particles.
Such effects do not directly affect the physics of our DM candidate
$X_-$, but they can affect the physics of the NLOP, which is a charged
and colored fermion.  Since coannihilations have to be taken into
account in our set-up, it is important to estimate the above effects.
We will not perform a detailed quantitative study of bound state
formation at finite temperature since, as we will argue below based on
qualitative estimates, such effects can most likely be neglected.  On
the other hand, certain higher order corrections, which take into
account the long-range Coulomb-like forces of QCD in the
deconfinement phase, are potentially relevant.  We find, however, that
their eventual contribution to the perturbative cross-sections is very
small.  Since strong interactions play the major role, we will neglect
in this subsection the effects due to electroweak interactions.

Let us first consider the issues associated with the formation of
bound states.  The potential problem is very simple.  If meta-stable,
$\Z_2$-even, bound states of two NLOP particles can form, their decay
through the self-annihilation of its constituents would lead to an
effective depletion of $\Z_2$-odd particles, consequently reducing the
DM relic density.  Bound state effects of DM colored particles
(gluinos) have been previously considered in \cite{Baer:1998pg}, in
connection with relic density computations, and were shown to have the
potential to reduce the final DM density by orders of magnitude
compared to the perturbative estimate.  The crucial difference here is
that the NLOP's are unstable, and as we argue below they decay into
LOP's well before they have time to form bound states.  The NLOP can
decay via inter-generational mixing into the LOP and a light quark
(decays into tops are forbidden by phase space in the quasi-degenerate
region of interest to us).  To lowest order in $\Delta$, as defined in
Eq.~(\ref{Delta}), the NLOP lifetime is
\be
\tau_{\psi} \approx \frac{8 \pi}{3 \lambda^2} \frac{m_\psi^{-1}}{\Delta^2}
\simeq \frac{ 2\times 10^{-23}}{\lambda^2} \, {\rm sec}~,
\label{psidecay}
\ee
where $\lambda^2 = \sum_{q=c,u} (g_{qL}^2+g_{qR}^2)$ is the
(model-dependent) coupling of the NLOP to the up and charm quarks, and
the number quoted is obtained by taking a rather degenerate case,
$m_{X_-} = 350$ GeV and $m_\psi=360$ GeV.

We estimate next the time scale when the (NLOP--NLOP) bound states
would become meta-stable.  A reasonable criterion is to assume that
these bound states are meta-stable against thermal fluctuations when
the temperature drops below their binding energy $E_{bind}$.  Since
the bound state system is non-relativistic, and in fact the scales are
such that the QCD interactions are in the perturbative regime, we can
estimate $E_{bind}$ in analogy to
positronium.~\footnote{\label{Debye}Strictly speaking, at finite
temperature there is a screening effect on charged particles, leading
to an effective Debye mass for the gluon, $m_g \simeq \sqrt{2} g_s(T)
T$.  It is easy to check, however, that for $T/m_\psi \simeq 10^{-2}$
[see Eq.~(\ref{bindingenergy})], the thermal screening can be
neglected.} The color Coulomb force between two quarks in a color
singlet state are obtained from the usual Coulomb potential by the
replacement $\alpha\rightarrow (4/3)\, \alpha_s(p)$.  Here
$\alpha_s(p)$ is evaluated at the typical momentum scale of the
virtual gluons responsible for the Coulomb-like interactions, which is
$p \sim m_{\psi} \alpha_s(p)/2$.  For $m_{\psi}\simeq 400$ GeV --the
typical scale in our scenario-- one gets $p\simeq 30$ GeV and
$\alpha_s(p) \simeq 0.14$.  The binding energy is then given by
\be
E_{bind}\simeq \frac{1}4\left(\frac{4}{3}\right)^2 \alpha_s(p)^2\,  
m_\psi \sim 10^{-2} m_\psi~.
\label{bindingenergy}
\ee
Thus, the (NLOP--NLOP) bound states become meta-stable \textit{after}
freeze out, which occurs when $T\sim m_\psi /25$.  In this radiation
dominated era, these two events happen at times $t\sim 10^{-7}$ sec.
and $t\sim 4 \times 10^{-9}$ sec.  in the evolution of the universe,
respectively.  It is therefore clear from Eq.~(\ref{psidecay}) that
for any reasonable value of the model-dependent coupling $\lambda$,
the NLOP's decay into LOP's immediately after freeze-out, and well
before bound states can become meta-stable.

The second effect we consider is due to the long-range Coulomb
interactions that, for sufficiently non-relativistic particles, can
distort their wave functions from the plane wave shape and alter the
standard quantum field theory perturbative computation of scattering processes.  Such
an effect is well-known and was first analyzed by Sommerfeld in the
QED context~\cite{Sommerfeld}.  It has recently been considered, in
connection with relic density computations, in \cite{Hisano:2006nn}
for electroweak interactions and in~\cite{Baer:1998pg} for strong
interactions.  For QED at zero temperature, the ``Sommerfeld'' effect
can be encoded in an effective parameter $S$ which reads, for
absolutely stable particles,\footnote{For unstable particles, one has
to check whether the Sommerfeld effect has time to take place, by
comparing the typical time scale of Coulomb interactions $t_{Coulomb}
\sim 1/(M v^2)$ with the decay time $\tau$ of the particle
\cite{Fadin:1988fn}.  If $\tau \ll t_{Coulomb}$, the Sommerfeld effect
has no time to take place.  In our case, $\tau_{NLOP} \gg t_{Coulomb}$
and thus we can effectively treat the initial NLOP particles as
absolutely stable.}
\be
S = -\frac{\pm x}{1-e^{\pm x}}\,,
\label{Sommerfeld}
\ee
with $x=\pi \alpha/v$, $v$ the velocity of the colliding particles in
the center of mass frame, and $\pm$ refer to repulsive or attractive
Coulomb forces, respectively.  Given a cross-section $\sigma$ computed
in the standard perturbative fashion between charged non-relativistic
initial states, the replacement $\sigma\rightarrow S\sigma $ is an
effective way to take into account the Coulomb forces, assuming
relativistic final states.  In a relativistic approach, the Sommerfeld
factor $S$ is obtained by resumming an infinite class of Feynman
diagrams (``ladder'' diagrams).  From Eq.~(\ref{Sommerfeld}), it is
clear that $S$ is non-negligible only for sufficiently
non-relativistic particles.  Cold DM candidates are by definition
non-relativistic at freeze-out and the Sommerfeld effect can play a
role, as emphasized in \cite{Baer:1998pg, Hisano:2006nn}.  For strong
interactions in a perturbative regime, an analysis along the lines of
QED can be made.  By neglecting screening effects due to temperature
(which is a good approximation at $T_f$, see also footnote
\ref{Debye}), Eq.~(\ref{Sommerfeld}) still holds, with the replacement
$\alpha\rightarrow C_r \alpha_s(p)$ in $x$, where $C_r$ is a color
factor that depends on the $SU(3)$ representation of the particles
involved, and $p \simeq m v$ is the typical momentum of the virtual
gluons responsible for the Sommerfeld effect in a relativistic
treatment \cite{Baer:1998pg}.  For a $\psi$-$\bar \psi$ pair in the
initial state, one can have either a singlet or an octet $SU(3)$
configuration, with color factors $C_{\bf 1} = 4/3$ and $C_{\bf 8} =
-1/6$, respectively At freeze-out, $v\simeq \sqrt{2T_f/m_\psi} \simeq
0.25$ and we get, for $m_\psi \simeq 400$ GeV, $S_{\bf 1}\simeq 2.3$
and $S_{\bf 8}\simeq 0.9$.  Thus, the Sommerfeld effect, especially in
the singlet channel, is non-negligible.  We have quantified the impact
of the Sommerfeld effect on the computation of $\Omega$ by replacing
the averaged tree-level cross-sections as follows:
\bea
\sigma_{\psi\bar\psi\rightarrow gluons} &\rightarrow&
\frac{1}{9} \left[ S_{\bf 1} \, 
\sigma_{(\psi\bar\psi)_{\bf 1}\rightarrow gluons}+ 
8 \, S_{\bf 8} \, \sigma_{(\psi\bar\psi)_{\bf 8}\rightarrow gluons} \right]~,
\nonumber \\
\sigma_{\psi\bar\psi\rightarrow q\bar q} &\rightarrow&
\frac{1}{9} \left[ S_{\bf 1} \, 
\sigma_{(\psi\bar\psi)_{\bf 1}\rightarrow q\bar q}+ 
8 \, S_{\bf 8} \, \sigma_{(\psi\bar\psi)_{\bf 8}\rightarrow q\bar q} \right]~. 
\nonumber
\eea
As it turns out, the enhancement of the cross-sections in the
attractive singlet channel is largely compensated by the repulsive
octet channel, so that the effect on the final $\Omega$ is always very
small (at most a few percent).

\subsection{Fine-tuning}
\label{sec:finetuning}

\begin{figure}[t]
\vspace*{-1.cm}
\centerline{ \hspace*{-0.5cm}
\includegraphics[width=0.467 \textwidth]{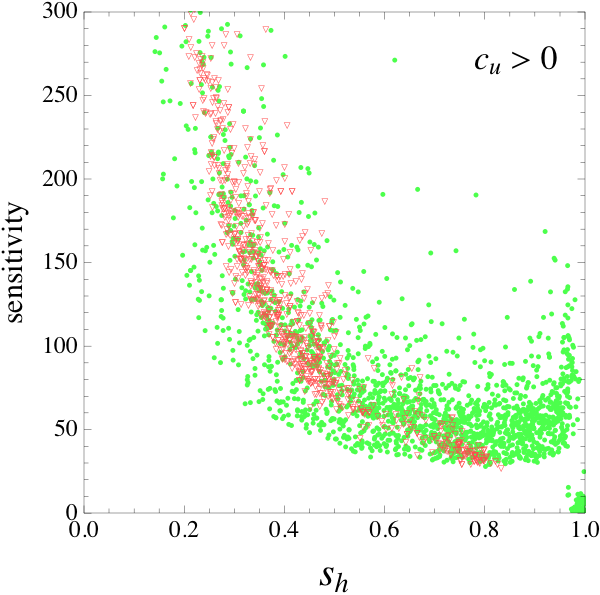}
\hspace*{0.5cm}
\includegraphics[width=0.45 \textwidth]{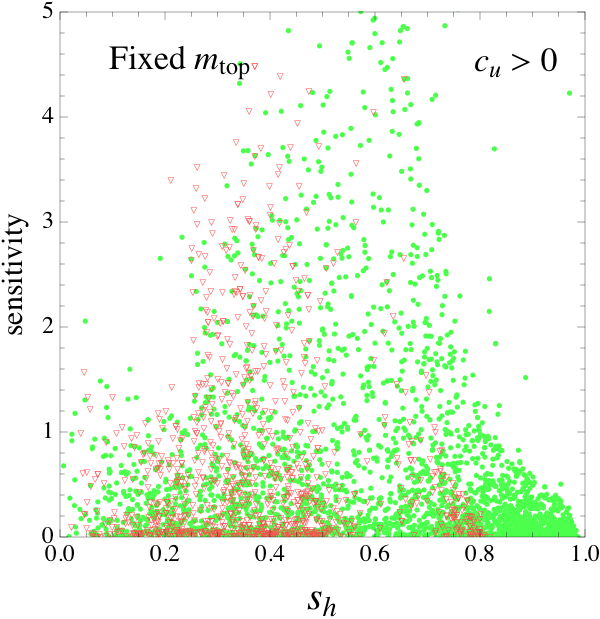}
}
\caption{Left panel: logarithmic sensitivity as a function of
$s_{h}$.  Right panel: logarithmic sensitivity with the requirement
that the top mass be kept fixed (see text).  The green dots
correspond to a random scan over parameter space with $c_{u} > 0$.
The subset that satisfies $140~{\rm GeV} < m_{t}(\mu \sim \mu_{IR}) <
160~{\rm GeV}$ is indicated by red triangles. The results for the
$c_{u} < 0$ region are similar.}
\label{Fig:finetuning}
\end{figure}

We established above that a rather well-defined region with
$|c_{u}| \sim 1/2$ is simultaneously selected by the minimization of
the Higgs potential, the EW constraints and the observed DM relic
density.  We should also recall that the pseudo-Goldstone nature of
the Higgs allows for its mass to be parametrically lower than the
scale of the KK resonances $\mu_\IR$, potentially alleviating the
little hierarchy problem present in other RS constructions (of course,
the large Planck-weak scale hierarchy is explained by the RS
mechanism).  It is therefore natural to ask how fine-tuned these
scenarios really are.  We quantify the fine-tuning at a given point in
parameter space by considering the sensitivity to the microscopic
parameters of the theory as measured by the logarithmic derivative
\cite{Barbieri:1987fn}
\be
{\textrm{sensitivity}} = {\rm max} 
\left\{ \left| \frac{\partial \log \langle h \rangle}{\partial \log \lambda_{i}} 
\right|
\right\}~,
\label{finetuning}
\ee
where $\lambda_{i} = c_{q_{1}}, c_{q_{2}}, c_{u}, c_{d}, m_{u}, M_{u},
m_{d}, M_{d}$ are the fundamental parameters of the model.  We find
that the apparent fine-tuning is dominated by the sensitivity to
$c_{u}$, as expected from our previous discussions.  We show in the
left panel of Fig.~\ref{Fig:finetuning} the sensitivity parameter for
the random scans described in section~\ref{sec:DMdensity}, as a
function of $s_{h}$.  These show that throughout the region of
parameter space that leads to phenomenologically viable EWSB breaking
minima, the apparent fine-tuning is worse than a percent.
Without tractable analytic expressions it is hard to identify
precisely the nature of such sensitivity, and it is conceivable that
the exponential nature of the warp factor gives rise to a generic
sensitivity.  In a 4D dual language, it would probably be related to
the usual generic sensitivity which affects theories with a dynamical
generation of scales, such as QCD or technicolor theories
\cite{Anderson:1994dz}.  It is then possible that the fine-tuning
estimate as given by Eq.~(\ref{finetuning}) is too conservative and a
more refined analysis is necessary.  It would be interesting to
investigate in more detail this issue.

However, we notice here that the sensitivity to the microscopic
parameters seems to be almost exclusively related to the requirement
of obtaining a large enough top Yukawa coupling.  In order to
illustrate the point, we compute again the logarithmic derivatives in
the $(c_u,M_u)$ plane, along directions with constant top mass.  The
result is shown in the right panel of Fig.~\ref{Fig:finetuning}.
Although we have not explored the logarithmic derivatives at constant
top mass in the whole parameter space and thus cannot exclude the
existence of other directions where the sensitivity is higher,
Fig.~\ref{Fig:finetuning} shows that part of the fine-tuning in the
model is associated with reproducing the observed top mass.  Similar
results also hold in the original MCHM$_{5}$ model without DM. We
therefore conclude that the present model necessitates some degree of
fine-tuning to reproduce the observed top mass, but once this
measurement has been done, the physical properties of the resulting
vacuum are fairly robust against variations in the microscopic
parameters of the theory (observables other than $\langle h \rangle$
also exhibit this property).  Notice that this is a single measurement
associated with the zero-mode sector, and therefore there is a sense
in which there is little sensitivity to the details of the new physics
beyond the standard model.  This also underscores the role that the
heavy top plays in driving EWSB.

The above situation should be contrasted with supersymmetric (SUSY)
scenarios that also present, generically, a sensitivity of order a
percent to various fundamental parameters.  In the SUSY case,
reproducing a large top mass is no more difficult than in the SM (it
is enough to choose the top Yukawa coupling, one of the
``fundamental'' parameters of these theories, to be of order one).
However, there is a well-known intrinsic fine-tuning associated with a
cancellation between the $\mu$ term and the soft SUSY breaking
parameters in the Higgs sector.  The latter depend quadratically on
the stop soft SUSY breaking masses, which in turn need to be taken
somewhat heavy (at least in the simplest SUSY extensions of the
standard model) in order to get a large enough Higgs quartic coupling
that allows satisfying the LEP bound on the Higgs mass.  If, in
addition, one takes into account the RG running from a high scale, and
quantifies the fine-tuning by the sensitivity to the
\textit{high-energy} parameters of the theory, the situation worsens.
In warped scenarios, the RS mechanism eliminates any possible
fine-tuning due to running from a high scale.  The dynamical
generation of the weak scale in gauge-Higgs unification scenarios,
allows for a further natural separation between the weak scale and the
KK scale $\mu_\IR$.  In fact, EWSB is generic, except that in most of
parameter space it is characterized by $s_{h} = 1$, which is not
phenomenologically acceptable.  The sensitivity shown in
Fig.~\ref{Fig:finetuning} is associated with the requirement $0 <
s_{h} < 1$.  Our observation is that all of this sensitivity is
actually associated with getting the correct top mass (which in these
scenarios is related in a relatively complicated manner to the
fundamental parameters of the model), rather than with the LEP bound
on the Higgs mass, which as Fig.~\ref{fig:mhmD} shows is easily
satisfied in these scenarios.

\section{Dark Matter in Other Models with Warped Extra Dimensions}
\label{sec:othermodels}

In the previous section, we showed how a very simple extension of a
minimal composite Higgs model in warped extra dimensions successfully
accounts for the observed DM relic abundance, while leading to a
non-trivial connection with the physics of EWSB and EW constraints. We now
briefly show that our prescription can be easily implemented in
essentially any model with warped extra dimensions.

\subsection{RS Model with Bulk Gauge and Fermion Fields 
\label{section:RS}}

This is the simplest model in which our mechanism can be applied.  We
will see that it is possible to get the correct DM relic density,
although some fine-tuning is likely required.  The bulk gauge symmetry
is taken to be $SU(3)_c\times SU(2)_L\times U(1)_{X_{1}}\times
U(1)_{X_{2}}$ with the Higgs field localized on or near the IR brane.
The SM $U(1)$ hypercharge factor is identified with the even linear
combination, $U(1)_{Y}=U(1)_{X_{+}}$.  The DM candidate $X_{-}$ is the
first KK mode of the 5D odd combination, Eq.~(\ref{Xminus}), and has a
mass given by Eq.~(\ref{DMmass}), $m_{X_{-}}\approx \sqrt{2/k L}\,
\mu_\IR \approx 0.24\, \mu_\IR$, where we have used $k L
\sim\log(M_P/\mathrm{TeV})\approx 34$.  Assuming no brane kinetic
terms, the IR scale in this model must obey $\mu_\IR\gtrsim 3$
TeV~\cite{Huber:2001gw} in order to be consistent with the EW
precision measurements.\footnote{Allowing for the Higgs to be a bulk
field with an exponential localization towards the IR
brane~\cite{Davoudiasl:2005uu} reduces this lower bound to
$\mu_\IR\gtrsim 2$ TeV. Similarly, sizable IR brane kinetic terms for
the gauge bosons can also decrease the corresponding bound on the
masses of the KK modes
significantly~\cite{Carena:2003fx}.\label{delocalized:Higgs}} The
annihilation rate of $X_{-}$ is again fixed by the 4D coupling
$\tilde{g}$ between $X_{-}$, the lightest $\Z_{2}$-odd fermion,
$\psi_-^{(1)}$, and the SM zero-mode field $\psi_+^{(0)}$, and their
masses.  A simple computation shows that $\tilde{g} \lesssim Y
g^\prime$, with $g^\prime$ and $Y$ the SM hypercharge coupling and
hypercharge quantum number of $\psi_+^{(0)}$.  The $\Z_{2}$-odd
fermions arise as explained in section~\ref{Z2} by doubling a subset
of the (SM) fermion fields.  These must be chosen to have a
localization parameter $|c| < 1/2$, so that the resulting $\Z_{2}$-odd
fields are heavier than $X_{-}$.  In scenarios that explain the
fermion mass hierarchies by fermion localization, this leaves as
natural candidates the quarks in the third generation.  The
self-annihilation rate $\sigma(X_{-}X_{-}\rightarrow \psi_+^{(0)}
\psi_+^{(0)})$ is too small to give the correct relic density $\Omega
h^2$, which is correspondingly too high.  However, the DM relic
density can be lowered to the observed level provided coannihilation
processes with the colored quarks $\psi_-^{(1)}$ are important.  This
requires $\psi_-^{(1)}$ and $X_{-}$ to be fairly degenerate, hence
$|c|$ to be very close (from below) to 1/2.  It turns out to be hard
to take $c_{q_L} \simeq 1/2$ or $c_{t_R}\simeq -1/2$, because in order
to get the correct 4D top Yukawa coupling one is forced to increase so
much the 5D top Yukawa coupling that it would enter the strong
coupling regime~\cite{Agashe:2003zs}.  The alternative is to double
the right-handed bottom quark only, and take $c_{b_R}\simeq -1/2$.  In
this case, other scattering processes become important, namely
$\sigma(X_{-}b_{R-}\rightarrow b_R g)$, $\sigma(b_{R-}\bar
b_{R-}\rightarrow g g)$ and $\sigma(b_{R-}\bar b_{R-}\rightarrow q\bar
q)$.  Given enough degeneracy between $X_{-}$ and $b_{R-}$, the relic
density of $X_{-}$ can match the observation.  For instance, for
$m_{X_{-}}=750~ (500)$ GeV, which corresponds to $\mu_{\rm IR}\simeq
3~(2)$ TeV (see footnote~\ref{delocalized:Higgs}), one requires
$c_{b_R}\simeq -0.496~(-0.495)$.  In terms of mass splittings, one
gets $\Delta \equiv (m_{b_{R-}}-m_{X_{-}})/m_{X_{-}}\simeq 6~(9)\%$.
This is not a fully satisfactory scenario due to the moderate tuning
needed (and the lack of independent motivation for the choice of
parameters), but it shows how this idea can be successfully
implemented even in the simplest 5D warped model.

By relaxing the explanation of hierarchical Yukawa couplings due to
the fermion localization one can envisage a different scenario, where
all (light) SM fermions share the same profile in the extra dimension.
Such a situation occurs in Higgsless models, which is the subject of
the next subsection.

\subsection{Higgsless Models}

Higgsless models induce EWSB by means of boundary conditions in the
extra dimension~\cite{Csaki:2003dt}.  In this way, the SM gauge bosons
acquire their longitudinal components through the Higgs mechanism but
no extra scalar degree of freedom (the Higgs) is present in the low
energy spectrum.  Unitarity violations in longitudinal gauge boson
scattering are delayed by the exchange of the gauge boson KK
excitations up to the cut-off of the theory.  Recently, using ideas
borrowed from~\cite{Agashe:2003zs,Agashe:2006at}, the first
5-dimensional Higgsless model roughly compatible with EW precision
data at tree level has been
presented~\cite{Cacciapaglia:2006gp}.~\footnote{Deconstructed versions
of Higgsless models are more flexible than 5-dimensional ones and
models compatible with data, at the tree level, have been
constructed~\cite{SekharChivukula:2006cg}.} A crucial ingredient is
the delocalization of the (LH) light
fermions~\cite{Cacciapaglia:2004rb}, that have a common localization
parameter close to the conformal point $c_L\lesssim 1/2$.  This means
that we can double the whole (LH) fermionic spectrum, following the
prescription described in section~\ref{Z2}, with the corresponding
increase in the number of open channels for the DM particle to
annihilate.  Furthermore, the low IR scale favors a stronger
annihilation cross-section and naturally produces the right order of
magnitude for the relic abundance.  Our starting point is the model of
Ref.~\cite{Cacciapaglia:2006gp}, in which $c_L=0.46$.  In this case,
the bulk gauge group is $SU(2)_L\times SU(2)_R \times U(1)_{X_{1}}
\times U(1)_{X_{2}}$, with $U(1)_{B-L}=U(1)_{X_{+}}$ the even
combination of $X_{1}$ and $X_{2}$.  The smaller warp factor and lower
IR scale in this model lead to a relatively light DM candidate, with
mass $m_{X_{-}}\sim 0.4 \,\mu_\IR \sim 114$ GeV. Doubling the whole LH
light spectrum leads to a too large annihilation cross section, mostly
due to the larger charge of the leptonic fields under the $U(1)_{B-L}$
group, whereas doubling only the light LH quarks results in a
cross-section that is a bit too small.  As an example, doubling only
one LH lepton doublet, we obtain $\Omega h^2 \sim 0.07$.  This is
impressively close to the observed value, considering that we have
taken identical values for all the parameters, as in
Ref.~\cite{Cacciapaglia:2006gp}.

\section{Phenomenology of the Dark Matter Sector}
\label{sec:phenomenology}

\subsection{Collider Phenomenology}

The collider signatures of our construction share common qualitative
features with other models of new physics with stable particles.  New
states are produced in pairs and (cascade) decay to SM particles and
the LOP, which is perceived as missing energy in the detector.  The
specific details of the spectrum of NLOP induce, however, significant
differences at the quantitative level.

In order to be more specific, we will discuss the phenomenology of the
GHU model described in section~\ref{sec:GHU}.  We study, as an
interesting and representative example, the case with $c_u>0$.  The
light $\Z_2-$odd spectrum consists of the LOP with a mass $m_{X_-}\sim
300-400$ GeV and the first KK modes of the bidoublet component of
$\xi_u^-$, that contain two quarks of charge $2/3$, one of charge
$5/3$ and one of charge $-1/3$.  We will denote these quarks as
$t^\prime$, $q^u$, $\chi^u$ and $q^d$, respectively, and, generically,
as $\psi$.  They are degenerate, with a mass $m_\psi-m_{X_-}\sim 0.15
\,m_{X_-}$, except for small EWSB effects that make one of the charge
$2/3$ quarks ($t^\prime$) slightly lighter.

Vector-like quarks are pair-produced at the LHC via the QCD
interactions with a model-independent cross-section that depends only
on the mass of the quark~\cite{pairproduction},
\begin{equation}
\sigma_{pp\to Q\bar{Q}} \sim 15 ~(1.5)~ \mathrm{pb},
\mbox{ for } m_Q\sim 400~
(600) \,\mathrm{GeV}.
\end{equation}
The distinctive feature is the high degree of degeneracy between the
NLOP and the LOP, which forbids the natural cascade decay through a
top quark, leaving decays into light jets and missing energy (the LOP)
as the main signature.  The charge $5/3$ quark cannot directly decay
into the LOP and SM particles, and therefore undertakes a cascade
decay through an off-shell $W^\ast$ to $t^\prime$, which then decays
to jets and missing energy.

The decay width for the two-body decays (kinematically allowed for
$q^d$ and, through inter-generational mixing for $t^\prime$ and $q^u$)
is given, to leading order in the mass difference, by (the inverse of)
Eq.~(\ref{psidecay})
\begin{equation}
\Gamma (\psi \to jX_- )
\approx
\frac{3 \lambda^2}{8\pi}
\Delta^2 m_\psi 
\approx
(4\times 10^{-6}\,\mathrm{GeV})
\left(\frac{\lambda}{0.002}\right)^2
\left(\frac{\Delta }{0.15}\right)^2
\frac{m_\psi}{460},
\label{two:body}
\end{equation}
where $j=u,c,b$, the relevant coupling is denoted with $\lambda$,
$\psi$ stands here for any of $t^\prime,q^u$ and $q^d$, and $m_\psi$
is measured in GeV. For the numerical result we have replaced the
different parameters with typical values for the charge $2/3$ quarks
(we have assumed inter-generational mixing in the up sector to be of
the order of the corresponding Cabbibo-Kobayashi-Maskawa matrix
elements $\lambda\sim V_{ub,cb} \tilde{g}/\sqrt{2} \lesssim 0.2 \times
(10^{-2}-10^{-3})$).  In the case of $q^d$ there is no
inter-generational mixing suppression ($\lambda=\tilde{g}$) and the
width can easily be
\begin{equation}
\Gamma(q^d \to bX_- ) \approx (10^{-1}-10^{-2})~ {\rm GeV}.
\end{equation}
The corresponding three-body decay for $\chi^u$ proceeds through an
off-shell $W^\ast$, which can decay leptonically.  In the contact
interaction approximation, the partial decay width into muons is
\begin{eqnarray}
\Gamma(\chi^u \to t^\prime \mu^+ \nu_\mu) 
&= &
\frac{G_F^2 m_{\chi^u}^5}{192 \pi^3} f(x)
\approx
(6\times 10^{-5}\, \mathrm{GeV})
\left(\frac{m_{\chi^u}}{500\,\mathrm{GeV}}\right)^5 \frac{f(x)}{f(0.9)}~,
\end{eqnarray}
where $f(x) = 1-8x^2 +8x^6 -x^8 -12x^4 \log x^2$, with 
$x\equiv m_{t^\prime}/m_{\chi^u}$.

Thus, we see that the NLOP are produced with a large cross section and
decay promptly, leading to the following typical signatures,
\begin{eqnarray}
pp&\to& t^\prime \bar{t}^\prime, q^u \bar{q}^u, q^d \bar{q}^d \to
 jj \not \hspace{-3.5pt}E_T,\\
pp&\to& \chi^u \bar{\chi}^u \to t^\prime \bar{t}^\prime W^\ast W^\ast 
\to l \nu jjjj  \not \hspace{-3.5pt} E_T. 
\end{eqnarray}
The former signature, that can benefit from a larger cross-section due
to the quark multiplicity, is challenging due to the lack of leptons
to trigger on, and the fact that the amount of transverse missing
energy is limited by the small mass difference between the NLOP and
the LOP. The latter, more promising due to the presence of a
leptonically decaying $W$, has however the problem of the extra source
of missing energy (the neutrino from the $W^\ast$) and the fact that
the $W^\ast$ is off its mass shell and therefore its mass cannot be
reconstructed.  A detailed analysis, which is beyond the scope of this
work, is needed to asses the trigger efficiency in accepting the
signal and the best strategy to search for this new sector in models
with warped extra dimensions, but the large production cross-sections
seem to indicate that discovery should be possible at the LHC.

Thus, although the generic collider implications of our construction
share some features with other models of new physics with stable
particles, the particular details of the spectrum of NLOP gives a very
characteristic signature, with direct decays into jets plus missing
energy and, in some cases, also short cascade decays through off-shell
$W$'s.  These are challenging signals at the LHC, but the large cross
sections due to the low mass of the new particles and the high
multiplicity should help in the discovery of these channels.  One
should also remember that the $\Z_2$-even sector of these models is
also predicted to have light resonances.  In the particular case we
are discussing, we have a light quark doublet with hypercharge $7/6$
that will be easily observed in the early phase of the
LHC~\cite{Contino:2008hi}.  The more challenging, although
possible~\cite{KKgauge}, discovery of the gauge boson KK excitations
and the lightest particles of the $\Z_2$-odd sector as we have
described above, should then draw a clear picture of the structure of
EWSB.

\subsection{Direct Dark Matter Searches}

Direct detection of DM particles is mostly based on detectors in which
the relevant process is the scattering amplitude
DM-atoms$\rightarrow$ DM-atoms.  Since, as we have seen, $X_-$
typically couples only to third generation quarks (the top quark in
the GHU model, $b_R$ in the RS model),
its direct detection rate is expected to be too small for current and
future planned experiments.  We can be more quantitative and estimate
the cross-section $X_-N\rightarrow X_-N$, where $N$ is a nucleon.  Not
surprisingly, the situation is analogous to that of UED where the DM
particle is identified with the first KK mode of the 5D hypercharge
gauge field, with the important difference that in our case there is
no Higgs exchange and only heavy quarks $Q$ are relevant.  Following
\cite{Cheng:2002ej} and using their notation, the spin-independent
cross-section reads
\be
\sigma_{scalar} = \frac{m_N^2}{4\pi (m_{X_-}+m_N)^2} f_{N}^2\,,
\ee
where $f_{N} = \beta_Q \langle N | \bar{Q} Q |N\rangle$ with $Q$ top or
bottom quarks, and
\be
\beta_Q \simeq m_Q (Q_{X,L}^2 \tilde g_L^2+Q_{X,R}^2 \tilde g_R^2)
\frac{m_{X_-}^2+m_\psi^2}{(m_{X_-}^2-m_\psi^2)^2}\,,
\label{betaQ}
\ee
with $m_\psi$ the mass of the NLOP. A careful estimate of the nuclear
matrix element $\langle N| Q\bar Q|N\rangle$ would require a detailed
one-loop analysis, along the lines of \cite{Drees:1993bu}.  For an
order of magnitude estimate, it is however enough to use the old
result \cite{Shifman:1978zn}
\be
\langle N| \bar{Q} Q|N\rangle  = \frac{2}{27} \frac{m_N}{m_Q} 
(1-\sum_{q=u,d,s} f_{T_q}^N)~,
\ee
where $f^N_{T_q} = \langle N| \bar{q} q | N \rangle m_q/m_N$.  It is
now straightforward to compute $\sigma_{scalar}$.  For the GHU model
we considered, by taking, say, $m_{X-}\simeq 350$ GeV, $m_\psi \simeq
380$ GeV, $\tilde g_L =0$, $\tilde g_R\simeq 0.25$, $Q_{X,R}=2/3$, one
has
\be
\sigma_{scalar} \simeq 2\times 10^{-10}\,  {\rm pb}\,,
\label{sigmascalar}
\ee
which is a value too low for current experiments and would require new
experiment proposals, such as super-CDMS \cite{Akerib:2006rr}.  Due to
the higher value of $m_{X_-}$, $\sigma_{scalar}$ is even smaller in
the RS model with doubled $b_R$ considered in
subsection~\ref{section:RS}.  In both models, due to the denominator
term in Eq.~(\ref{betaQ}), $\sigma_{scalar}$ can become sizable only
for extreme degenerate cases in which $m_{X_-}\simeq m_\psi$.

In Higgsless model, the situation is different, since one might also
have direct couplings of $X_-$ with light quarks.  In addition,
$X_{-}$ is lighter, $m_{X_-}\sim 100$ GeV, and hence DM direct
detection seems more promising.  However, in the specific model of
\cite{Cacciapaglia:2006gp}, $\sigma_{scalar}$ is still suppressed due
to the low values of the charges, $Q_{X,L}=1/6$ ($Q_{X,R}=0$),
resulting in a cross-section of the same order of magnitude as
(\ref{sigmascalar}).

\section{A Comment on Anomalies\label{sect:anomalies}}

We have assumed so far that the $\Z_2$ exchange symmetry is an exact
symmetry of the theory, namely that no quantum corrections can
possibly violate it.  
We show here that the exchange symmetry is exact 
by noticing that the CS terms that are required by gauge invariance
are always $\Z_2$-even and
hence invariant.  This is actually expected, since the $\Z_2$ symmetry
is a global symmetry which has nothing to do with parity, the discrete
symmetry typically broken by anomalies.  For simplicity, we will focus
our attention on the simplest RS scenario with fermion and gauge bulk
fields analyzed in subsection~\ref{section:RS}, but the main results
remain valid also for the more refined GHU model of
section~\ref{sec:GHU}.

The bulk gauge group is $G=SU(3)_c\times SU(2)_L \times
U(1)_{X_{1}}\times U(1)_{X_{2}}$, broken to $SU(3)_c\times SU(2)_L
\times U(1)_{Y}$ at the IR brane and fully unbroken at the UV brane.
The 5D fermion spectrum consists of massive Dirac fermions, one for
each SM fermion, with $(++)$ or $(--)$ b.c., depending on the
chirality of the SM fermion.  They have charge
$(\frac{1}{2}Y,\frac{1}{2}Y)$ under $U(1)_{X_{1}}\times U(1)_{X_{2}}$,
with $Y$ the corresponding SM hypercharge for the given fermion field,
while the 5D bottom fermion fields $\psi_{b_{1}}$ and $\psi_{b_{2}}$
have charges $(-1/3,0)$ and $(0,-1/3)$.  Although the 4D massless
fermion spectrum of the model is anomaly free by construction, being
identical to the SM spectrum, the 5D theory needs CS terms to restore
5D gauge invariance fully, because localized anomalies (globally
vanishing in 4D once integrated over the internal space) appear at the
UV and IR branes (see \cite{Scrucca:2004jn} for a review).\footnote{In
this simple example, localized anomalies appear only due to the
doubling we have performed.  The original model is fully anomaly free
since the SM anomaly cancellation applies also to the localized
terms.} Recall that the warping does not alter the localization
pattern of anomalies, which is then as in flat space
\cite{Hirayama:2003kk}.  It is extremely useful to also recall that
the form of a localized anomaly at a boundary is fully determined by
the ``effective'' chiral spectrum which is found there by neglecting
the b.c. at the other end-point, which is equivalent to sending the
other boundary to infinity (see e.g. \cite{Scrucca:2004jn} for a
derivation of this result).  Let us consider the UV brane, where both
$U(1)$ factors are unbroken, and focus on the possible anomalies
involving the gauge field $X_-$.  All SM fermions are neutral under
$U(1)_{X_{-}}$, with the exception of $\psi_{b_{1}}$ and
$\psi_{b_{2}}$, which have opposite charges $-1/3$ and $+1/3$.
Correspondingly, all $U(1)_{X_{-}}^3$, $U(1)_{X_{-}} U(1)_{Y}^2$,
$SU(3)_c U(1)_{X_{-}}$ and mixed $U(1)_{X_{-}}$--gravitational
anomalies trivially vanish at the UV brane.  At the IR brane, no
anomalies involving $U(1)_{X_{-}}$ can appear, since the gauge field
$X_-$ vanishes there.  It is not difficult to compute all other
localized anomalies.  One finds that the ``doubling'' procedure
induces pure $SU(3)_c$, $U(1)_{Y}^3$ and mixed $SU(3)_c U(1)_{Y}$,
$U(1)_{Y}$--gravitational anomalies at the UV and IR brane, as well as
a $U(1)_{Y} U(1)_{X_{-}}^2$ anomaly at the UV brane.  ll these anomalies are cancelled by suitable 5D CS terms, which involve either
0 (the former) or 2 (the latter) gauge fields $X_-$.~\footnote{Possible modifications to the
wave functions of gauge fields induced by such CS terms can be safely neglected, being one-loop suppressed.}  Consequently,
all CS terms are $\Z_2$-even, as anticipated.  A similar result is
found for the GHU model of section~\ref{sec:GHU}.

\section{Discussion and Conclusions\label{sect:conclusions}}

In this paper we have proposed a generic construction that allows to
endow given models with a DM candidate.  This is achieved through an
extension in which the model acquires a $\Z_{2}$ exchange symmetry.
The lightest $\Z_{2}$-odd particle is then absolutely stable.  

We have considered several models with warped extra dimensions.
Although these scenarios are well-motivated extensions of the SM,
explaining both the Planck-weak scale hierarchy as well as the flavor
structure of the SM, they do not contain, generically, stable
particles that can account for the observed DM component of the
universe.  We have shown that our mechanism can easily solve this
deficiency.

We have payed special attention to a class of warped scenarios that is
particularly appealing: GHU/composite Higgs scenarios.  In this case,
the $\Z_{2}$ structure responsible for the stability of the DM
candidate is tightly connected to the physics that leads to the
dynamical breaking of the EW symmetry.  As in supersymmetry, the dark
matter mass and couplings are intimately connected to the EW scale.
In fact, not only does our construction not introduce new parameters,
but there is a further sense in which it can be considered minimal.
As was emphasized in the main text, the physics of EWSB in such models
crucially depends on certain fields that have two properties: they are
fermionic fields without zero-modes and, through their strong
connection to the top sector, they give a significant --in fact,
crucial-- contribution to the Higgs potential.  This is precisely the
sector that gives rise to the $\Z_{2}$-odd particles that can lead to
a realistic dark matter candidate.  As a result, the $\Z_{2}$-odd
sector, through its contribution to the Higgs potential, plays a key
role in the realization of an EWSB vacuum with the desired physical
properties.  As explained in detail in the main text, the effects due
to the $\Z_{2}$-odd sector go in the direction of alleviating the
constraints imposed by low-energy measurements.  The picture that
emerges is rather compelling: the observed DM relic abundance and the
very precise measurements of the EW observables at LEP and SLC point
to a \textit{common} region in parameter space.  This region is
characterized by several fermionic resonances nearly degenerate with
the DM candidate (a spin-1 particle), which in turn is predicted to
have a mass in the $300-500~{\rm GeV}$ range.

We have also pointed out that although the GHU scenarios (with or
without DM) present a sensitivity of order one percent or worse with
respect to microscopic parameters (larger than what naive
considerations would indicate), this sensitivity disappears
once the top mass measurement is imposed.  Thus, the moderate
sensitivity of order a percent, that is present in virtually every
extension of the SM that has any relation to the physics of EWSB
acquires a new twist: it indicates an intrinsic difficulty in
accommodating a heavy top.  By contrast, once the top mass is fixed to
a value of order the EW scale, the Higgs mass easily satisfies the LEP
bound, while typically lying below $170~{\rm GeV}$ or so.

Regarding the phenomenological implications of our construction, the
low scale predicted for the new $\Z_2-$odd sector and, in particular,
new vector-like quarks, guarantees large production cross-sections.
Due to the particular features of the NLOP spectrum, which is very
degenerate with the LOP, the most common signature will be jets plus
missing energy accompanied in some cases by the semileptonic decays of
a pair of off-shell $W^\ast$.  These are challenging signatures that
will require dedicated analysis.  The large production cross-sections
and the information coming from the easier channels in the $\Z_2-$even
sector should however be sufficient to guarantee a discovery of the DM
sector at colliders.  We have also seen that DM direct detection is
not very promising in present or near future experiments, due to the
fact that couplings to light valence quarks are suppressed by
intergenerational mixing, leading to very small cross-sections.

\vspace{5mm} 
\noindent 
%

\section*{Acknowledgements}

We thank P. Batra, G. Cacciapaglia, B. Dobrescu, J. Hubisz, R. Iengo, K.C. Kong,
J. Lykken, A. Romanino, V. S. Rychkov and P. Ullio for useful
discussions.  Work partially supported by the European Community's
Human Potential Programme under contract MRTN-CT-2004-005104.  Work of
G.P. was partially supported by the European Union 6th framework
program MRTN-CT-2004-503069 ``Quest for unification'',
MRTN-CT-2004-005104 ``ForcesUniverse'', MRTN-CT-2006-035863
``UniverseNet'' and SFB-Transregio 33 ``The Dark Universe'' by
Deutsche Forschungsgemeinschaft (DFG).  E.P. was supported by DOE
under grant No.  DE-FG02-92ER-40699.  J.S. was partially supported by
US DOE under contract No.  DE-AC02-07CH11359 and by SNSF under
contract 200021-117873.  M.S. thanks Fermilab and J.S. the Aspen
Center for Physics for hospitality during the early stages of this
work.

\appendix

\section{The MCHM$_{5}$ with Dark Matter}
\label{App:Model}

In this section we summarize the model of GHU that was analyzed in the
main text.  The model has an $SO(5)\times U(1)_{X_{1}} \times
U(1)_{X_{2}}$ bulk gauge symmetry, with a discrete symmetry under
which the two $U(1)$ factors are exchanged.  The $\Z_{2}$-even sector
is identical to the MCHM$_{5}$ of Ref.~\cite{Contino:2006qr}.  The
$SO(5)/SO(4)$ directions are broken on both branes by choosing $(-,-)$
boundary conditions for the $\mu$ components
\begin{equation}
A_\mu^{\hat{a}}~(-,-)~,
\end{equation}
where $\hat{a}=1,\ldots,4$ runs over the $SO(5)/SO(4)$ indices.  The
corresponding components along the extra dimension, $A_5^{\hat{a}}$,
(which are four dimensional scalars) have zero modes that transforms
as a bidoublet of $SO(4)\sim SU(2)_L \times SU(2)_R$ and are
identified as the SM Higgs doublet.  The $\mu$ components of the
remaining gauge fields have the following b.c.:
\begin{eqnarray}
W^a_L & \sim & (+,+), \quad B \sim (+,+)~, \\
W^b_R & \sim & (-,+), \quad Z^\prime \sim (-,+)~,
\end{eqnarray}
where $a=1,2,3$, $b=1,2$, the $L,R$ indices correspond to the
$SU(2)_L\times SU(2)_R$ decomposition of $SO(4)$, and
\begin{equation}
B=\frac{g_{5X} W^3_R + g_5 X_{+}}{\sqrt{g_5^2+g_{5X}^2}}~,
\quad
Z^\prime=\frac{g_{5} W^3_R - g_{5X} X_{+}}{\sqrt{g_5^2+g_{5X}^2}}~.
\end{equation}
Here we have defined $X_{\pm} = (X_{1} \pm X_{2})/\sqrt{2}$, and
$X_{-\, \mu}$ satisfies $(+,-)$ b.c. Also, $g_{5X_1} = g_{5X_2} =
\sqrt{2} g_{5X}$.

The SM quarks are embedded in bulk fermions transforming in the
fundamental representation of $SO(5)$, $\mathbf{5}=(2,2)\oplus (1,1)$ with
$X_{+}$ charge $2/3$ and $-1/3$ for the up and down sectors,
respectively.  The odd fields couple as in
Eq.~(\ref{Aminuscouplings}).  The b.c. are as follows:
\begin{eqnarray}
\begin{array}{c}
\begin{matrix}
\xi_{q_1}=
\begin{bmatrix}
(2,2)^{q_1}_L =
\begin{bmatrix}q^\prime_{1L}(-+) \\ q_{1L}(++) \end{bmatrix}
&
(2,2)^{q_1}_R =
\begin{bmatrix}q^\prime_{1R}(+-) \\ q_{1R}(--) \end{bmatrix}
\\
(1,1)^{q_1}_L(-,-)
&
(1,1)^{q_1}_R(+,+)
\end{bmatrix},
\\  \\
\begin{array}{cc}
\hspace{-2.5cm} \xi^+_u=
\begin{bmatrix}
(2,2)^u_L(+-)
&
(2,2)^u_R(-+)
\\
(1,1)^u_L(-+)
&
(1,1)^u_R(+-)
\end{bmatrix},
&
\hspace{-2.5cm}
\xi^-_u=
\begin{bmatrix}
(2,2)^{u^-}_L(+-)
&
(2,2)^{u^-}_R(-+)
\\
(1,1)^{u^-}_L(-+)
&
(1,1)^{u^-}_R(+-)
\end{bmatrix},
\\ \\
\xi_{q_2}=
\begin{bmatrix}
(2,2)^{q_2}_L =
\begin{bmatrix}q_{2L}(++) \\ q^\prime_{2L}(-+) \end{bmatrix}
&
(2,2)^{q_2}_R =
\begin{bmatrix}q_{2R}(--) \\ q^\prime_{2R}(+-) \end{bmatrix}
\\
(1,1)^{q_2}_L(-,-)
&
(1,1)^{q_2}_R(+,+)
\end{bmatrix},
&
\xi_d=
\begin{bmatrix}
(2,2)^d_L(+-)
&
(2,2)^d_R(-+)
\\
(1,1)^d_L(-+)
&
(1,1)^d_R(+-)
\end{bmatrix}.
\end{array}
\end{matrix}
\end{array}
\label{fieldcomponents}
\end{eqnarray}
We have displayed the field content according to their $SO(4)$
decomposition, and $+$ and $-$ represent, respectively, Neumann or
Dirichlet boundary conditions at the corresponding brane.  Note that
the choice of parities above seems to allow for two SM doublet zero
modes per generation, coming from $q_{1L}$ and $q_{2L}$.  It is
actually only the symmetric combination $(q_{1L}+q_{2L})/\sqrt{2}$
that has a zero mode, the odd combination being coupled to a
UV--localized chiral fermion with a large mass.  The $O(4)\times
U(1)_X$ symmetry at the IR brane allows for the following mass mixing
terms:
\begin{equation}
m_u \overline{(2,2)}^{q_1}_L (2,2)^u_R
+ M_u \overline{(1,1)}^{q_1}_R (1,1)^u_L
+ m_d \overline{(2,2)}^{q_2}_L (2,2)^d_R
+ M_d \overline{(1,1)}^{q_2}_R (1,1)^d_L
+\mathrm{h.c.}
\end{equation}
Notice that the localized mass terms do not involve the odd fields in
$\xi^-_u$.  Thus, neglecting inter-generational mixing, we have a
total of eight parameters per quark generation, four bulk masses,
denoted in units of the bulk curvature $k$ by $c_{q_1}$, $c_{q_2}$,
$c_u$ and $c_d$ (the exchange symmetry forces $c_{u^-} = c_{u}$), and
four mass terms $m_{u,d}$, $M_{u,d}$.  The
first two generations have a negligible effect on EWSB and, for
the third generation, as long as $c_{q_1}<1/2$, $c_{q_1} < c_{q_2}$,
the two multiplets related to the bottom ($\xi_{q_2},\xi_d$) do not
play any significant role either.  We are therefore left with four
relevant parameters, $c_{q_1}$, $c_u$, $m_u$ and $M_u$.  Note however
that there is a phenomenological constraint on the value of $c_d$.
The reason is that, due to the chosen boundary conditions, the
component $(1,1)^d$ becomes ultralight for $c_d \lesssim
-1/2$~\cite{Agashe:2004ci}, easily violating limits from direct
production at the Tevatron~\cite{qprime:limits}.

\section{The Higgs Potential}
\label{HiggPotential}

The one-loop Higgs potential in 5D warped models does not admit a
simple analytic expression, which is related to the difficulty in
deriving an explicit form for the KK mass spectrum and Higgs
interactions for the relevant fields.  A useful tool to derive at
least an implicit but relatively compact form for the potential is
achieved by using the gauge-fixing outlined in \cite{Panico:2007qd} in
an holographic approach \cite{Luty:2003vm}.  In this way, the Higgs
potential is simply obtained by a rotation of the 4D holographic
fields only (see \cite{Panico:2007qd} for details).  The Higgs
potential in our model is actually the same as the one considered in
\cite{Contino:2006qr}, the only difference being the presence of the
$\Z_2$-odd SO(5) multiplet $\xi_u^-$.  The latter field, interestingly
enough, leads to a Higgs contribution which has opposite sign with
respect to that of 5D fermions admitting chiral zero modes.  Like
gauge fields, they contribute positively to the Higgs potential,
pushing the minimum towards zero.  In order to simplify the
discussion, let us consider a simple 5D model with SU(2) gauge
symmetry broken to U(1) by boundary conditions.  The contribution to
the ``Higgs'' potential given by two 5D fermion doublets $\Psi_{\pm}$
whose left-handed components satisfy the boundary conditions
\be
\Psi_L^+ = \left(\begin{array}{c}\psi_L^u (++) \\ \psi_L^d (--) \end{array}\right)\,, \ \ \ \
\Psi_L^- = \left(\begin{array}{c}\psi_L^u (+-) \\ \psi_L^d (-+) \end{array}\right)\,,
\ee
modulo an irrelevant constant term,  can be written as
\be
V_\pm(h) = -2 \int \frac{d^4p}{(2\pi)^4} \log\Big[1\pm c_{2h} \Pi(p,m_\pm)\Big]\,.
\label{app1}
\ee
In Eq.~(\ref{app1}), $\pm$ stands for the contributions of $\Psi^+$
and $\Psi^-$, respectively, $p$ is an Euclidean momentum and $c_{2h} =
\cos(2h/f_h)$.  All the non-trivial information on the mass spectrum
is encoded in the form factor $\Pi(p,m_\pm)$, with $p=\sqrt{p_\mu
p^\mu}$ and $m_\pm$ the bulk 5D mass terms.  If $|\Pi(p,m_\pm)|< 1$
over the whole integration region, then one can expand the log term in
Eq.~(\ref{app1}), which clearly shows the opposite contribution to the
potential given by the two 5D fermion fields.  One has
\be
\Pi(p,m) = \bigg[\frac{G_{--}(c)}{G_{-+}(c)} -\frac{G_{+-}(c)}{G_{++}(c)} \bigg]
\bigg[\frac{G_{--}(c)}{G_{-+}(c)} +\frac{G_{+-}(c)}{G_{++}(c)} \bigg]^{-1}\,,
\label{app3}
\ee
where $G_{\pm,\pm}(c)=G_{\pm,\pm}(c, p, z_{\IR}, z_{\UV})$ are
combination of the Bessel functions $J$ and $Y$:
\be
G_{\eta, \eta^\prime}(c, p, z_1, z_2)  = J_{c+\eta 1/2}(i \,p \, z_1) Y_{c+\eta^\prime 1/2}(i\, p\, z_2)-Y_{c+\eta 1/2}(i\, p\, z_1) J_{c+\eta^\prime 1/2}(i\, p \, z_2)\,,
\label{app4}
\ee
where $\eta,\eta^\prime = \pm$, and $c=z_\UV m$ is the usual
dimensionless mass term in warped space.  It is not difficult to check
that $|\Pi(p,m_\pm)|\leq 1$, it exponentially vanishes when
$p\rightarrow\infty$ and reaches its maximum value 1 just at the
origin $p=0$.\footnote{A similar result holds in flat space where
$\Pi(p,m) = -(p^2+m^2)/(m^2+p^2 \cosh(2 L \sqrt{p^2+m^2}))$, with $L$
being the length of the segment.  }

Let us now turn to the actual Higgs potential in our model, which is
complicated by the presence of localized IR mass terms and a
non-trivial gauge symmetry breaking pattern.  The most relevant fields
contributing to the potential are the gauge fields and the SO(5)
multiplets $\xi_u^\pm$, $\xi_d$, $\xi_{q_1}$ and $\xi_{q_2}$.  In the
holographic approach, these contributions are encoded in the
holographic gauge fields $W$ and $Z$, the bottom and top quarks, and
by a holographic $\Z_2$-odd fermion component of $\xi_u^-$.

The gauge field contribution reads (in the absence of any localized
IR or UV terms)
\be
V_g(h) = \frac{3}{2} \int \frac{d^4p}{(2\pi)^4} \bigg[2  \log\Big(1+s_{h}^2\frac{ \Pi_--\Pi_+}{2\Pi_+}\Big)+
\log\Big(1+s_{h}^2\frac{ \Pi_--\Pi_+}{2\Pi_+}\frac{2}{(1+\cos 2\theta_W)}\Big) \bigg]\,,
\label{app5}
\ee
where $\theta_W$ is the SM weak mixing--angle and
\be
 \Pi_\pm =  \frac{G_{\mp-}(1/2)}{G_{\mp+}(1/2)}\,,
\label{app6}
\ee
in terms of the functions defined in Eq.~(\ref{app4}).  In
Eq.~(\ref{app5}), the first and second log terms correspond to the
holographic $W$ and $Z$ contributions, respectively.

Let us now consider the fermion contribution to the Higgs potential.
Since by symmetry no IR mass terms can be introduced for $\xi_u^-$,
whose components satisfy boundary conditions of the $(+ -)$/$(- +)$
type, its contribution to the potential is given by $V_-(h)$ in
Eq.~(\ref{app1}), which can be rewritten (again, modulo constant
terms) as
\be
V_f^-(h) =  -2 N_c \int \frac{d^4p}{(2\pi)^4} \log\Big[1+s_{h}^2 \frac{\Pi_1}{2\Pi_0}\Big]\,,
\label{app7}
\ee
where $N_c=3$ is the QCD color factor and
\bea
\Pi_0 & = & - \frac{G_{-+}(c_u)}{G_{--}(c_u)}\,, \nn \\
\Pi_1 & = & -2\bigg(- \frac{G_{-+}(c_u)}{G_{--}(c_u)}+\frac{G_{++}(c_u)}{G_{+-}(c_u)}\bigg)\,.
\label{app7a}
\eea
The holographic bottom and top quark contributions are more involved.
The top contribution can be written as
\be
V_{top}(h) =    -2 N_c \int \frac{d^4p}{(2\pi)^4} \log\bigg[\Big(1+s_h^2 \frac{\Pi_1^u}{2\Pi_0^u}\Big)
\Big(1+s_h^2 \frac{\Pi_1^{q_1}}{2\Pi_0^q}\Big)-s_h^2 c_h^2\frac{ (M_1^u)^2}{2\Pi_0^u\Pi_0^q}\bigg]\,,
\label{app8}
\ee
in terms of the following form factors:
\bea
\Pi_0^u & = &  -  \frac{N_{q_1,u}^+(1/M_u)}{D_{q_1,u}(1/M_u)}\,, \nn \\
\Pi_1^u & = &  -2\bigg[- \frac{N_{q_1,u}^+(1/M_u)}{D_{q_1,u}(1/M_u)}+\frac{N_{q_1,u}^+(m_u)}{D_{q_1,u}(m_u)} \bigg]\,, \nn \\
\Pi_0^q & = &  \frac{N_{q_1,u}^-(m_u)}{D_{q_1,u}(m_u)} + \frac{N_{q_2,d}^-(m_d)}{D_{q_2,d}(m_d)}\,, \nn \\
\Pi_1^{q_1} & = & - \frac{N_{q_1,u}^-(m_u)}{D_{q_1,u}(m_u)} + \frac{N_{q_1,u}^-(1/M_u)}{D_{q_1,u}(1/M_u)}\,, \nn \\
M_1^u & = & \frac{4}{\pi^2 p^2 z_\IR z_\UV}\Big(\frac{m_u}{ D_{q_1,u}(m_u)} - \frac{1/M_u}{D_{q_1,u}(1/M_u)}\Big)\,,
\label{app9}
\eea
with
\bea
N_{i,j}^\pm(M) & = & G_{-\pm}(c_i) G_{+\pm}(c_j) +M^2 G_{+\pm}(c_i) G_{-\pm}(c_j)\,, \nn \\
D_{i,j}(M) & = &  G_{-+}(c_i) G_{+-}(c_j) +M^2 G_{++}(c_i) G_{--}(c_j)\,.
\label{app10}
\eea
In the above formulae, $m_u$ and $M_u$ are dimensionless IR brane mass
terms (see appendix~\ref{App:Model}).  The bottom quark contribution
is also given by Eq.~(\ref{app8}), provided one makes the
substitutions $c_{q_1}\leftrightarrow c_{q_2}$, $c_{u}\leftrightarrow
c_{d}$, $m_u\leftrightarrow m_d$, $M_u\leftrightarrow M_d$ in
Eqs.~(\ref{app8}) and (\ref{app9}).

It is useful to consider various limits for the IR brane mass terms.
When $m_u\rightarrow 0$, $M_u\rightarrow 0$, the form factor $M_1^u$
vanishes and the potential splits in the two contributions given by
$\xi_u^+$ and $\xi_{q_1,q_2}$.  As expected, in this limit the
contribution given by $\xi_u^+$ reduces to that of $\xi_u^-$, with
$\Pi_0^u\rightarrow \Pi_0$, $\Pi_1^u\rightarrow \Pi_1$, with $\Pi_0$
and $\Pi_1$ given in Eq.~(\ref{app7a}).  In the opposite limit,
$m_u\rightarrow \infty$, $M_u\rightarrow \infty$, the contribution
given by $\xi_u^+$ turns into that of a fermion with $(++)$/$(- -)$
boundary conditions, and is given by $V_f^+(h)$ in Eq.~(\ref{app1}).
This is again expected, since the large IR brane mass terms
effectively change the boundary conditions of the fermions on the IR
brane.  In the limits in which $M_u\rightarrow \infty$,
$m_u\rightarrow 0$ or viceversa, $\Pi_1^u\rightarrow 0$,
$M_1^u\rightarrow 0$, and hence $\xi_u^+$ does not contribute at all
to the potential.  As far as $\xi_{q_1,q_2}$ are concerned, one should
recall that bottom and top quarks are distributed in both multiplets
$\xi_{q_1}$ and $\xi_{q_2}$ and that a localized UV right-handed fermion
doublet is necessary to get rid of an unwanted zero mode left--handed
doublet \cite{Contino:2006qr}.  Such a localized field implies that
the contributions of $\xi_{q_1}$ and $\xi_{q_2}$ are entangled also in
the limit of vanishing or very large IR brane terms.  That is why
$\Pi_0^q$ in Eq.~(\ref{app9}) depends on both $c_{q_1,q_2}$ , $c_{u,d}$
and $m_{u,d}$.  Notice, however, that even in this case, if
$M_u\rightarrow \infty$, $m_u\rightarrow 0$ or viceversa, $\Pi_1^{q_1}$
vanishes, so that the whole top contribution to the potential vanishes
in this limit!  Such result is particularly clear in the approach
followed here, where the Higgs dependence of the holographic
Lagrangian is obtained by a SO(5) rotation of the 4D holographic
fields.  The rotation gives rise to Higgs--dependent terms when SO(5)
is broken at the IR brane by boundary conditions.  In the limit
$M_u\rightarrow \infty$, $m_u\rightarrow 0$ or viceversa, the singlet
and the bidoublet components of the SO(5) multiplets $\xi_u^+$,
$\xi_{q_1}$ and $\xi_{q_2}$ have the same boundary conditions at the IR
brane, implying SO(5) invariant boundary conditions and hence no
couplings of such fermions with the Higgs.  Actually, there is a whole
one--dimensional family of boundary conditions, when $m_u = 1/M_u$,
for which $\Pi_1^u=\Pi_1^{q_1}=M_1^u=0$, as is clear from
Eq.~(\ref{app9}).  Needless to say, similar results hold when
considering the IR brane terms $m_d$ and $M_d$.

The total one-loop Higgs potential is finally given by
\be
V_{Tot}(h) = V_g(h) + V_{top}(h) + V_{bottom}(h) + V_f^-(h) \,.
\label{app11}
\ee
%

\section{The Trilinear Couplings}
\label{coupling}

The holographic approach, used in the previous appendix to derive the
effective Higgs potential, can also be efficiently applied to the
computation of the trilinear couplings $\tilde g_L$ and $\tilde g_R$
of the DM candidate $X_-$ with the third generation quarks and the
$\Z_2$-odd fermions.  As explained in the main text, these couplings
are an essential ingredient to compute the DM relic abundance.

For simplicity, we neglect EWSB effects, which introduce only a slight
change ($\lesssim \textrm{few}\, \%$) in the computation of the DM
relic abundance, an accuracy that we have no interest in achieving.
Starting from Eqs.~(\ref{gRcoupling}) and (\ref{gLcoupling}), our aim
is to extract the trilinear couplings $\tilde g_R \bar t_R
\hspace{-2pt}\not \hspace{-2.7pt} X_- \psi_{-,R}$ (for $c_u<0$) and
$\tilde g_L \bar t_L \hspace{-2pt}\not \hspace{-2.7pt} X_- \psi_{-,L}$
(for $c_u>0$) where $\psi_{-,R}$ is the right-handed component of the
NLOP singlet state and $\psi_{-,L}$ is the left-handed component of
the NLOP field in the bidoublet coupled to the top.  In the
holographic approach, $\tilde g_{L/R}$ are obtained by writing the
bulk-to-boundary propagators for the holographic fields $t_{L/R}$,
$\psi_{-,L/R}$ and $X_-$ and integrating the vertices
(\ref{gRcoupling}) and (\ref{gLcoupling}) over the compact space,
taking care of normalization factors required to define canonically
normalized fields.  We can consider the cases $c_u>0$ and $c_u<0$ at
the same time.  The $(t_L, t_R)$ system can be described by choosing
as holographic degrees of freedom the $UV$ values of the LH components
of the $\xi_{q_1}$ (and $\xi_{q_2}$) multiplet and the RH ones of the
$\xi_u^+$ field (for $m_u\neq 0$).  In this way we obtain the
quadratic holographic Lagrangian,
\be
{\cal L} = \overline{t}_L \frac{\pslash}{p} \frac{\Pi_0^q}{z_{\UV}} \, t_L
+ \overline{t}_R \frac{\pslash}{p} \frac{\Pi_0^u}{z_{\UV}} \, t_R\,,
\ee
with $p=\sqrt{p_\mu p^\mu}$ and $p_\mu$ a Minkowskian momentum and the
form factors $\Pi_0^q$ and $\Pi_0^u$ are as in Eq.~(\ref{app9}) but
with $p\rightarrow -ip$.\footnote{The factor $1/z_{\UV}$ ensures that
the holographic fields have canonical dimension in $4D$.}

The holographic fields are not mass eigenstates but rather a
superposition of all the states in the $4D$ KK tower.  As a
consequence, we need some care to extract the couplings of the mass
eigenstates from the holographic Lagrangian.  To do this we use the
fact that the holographic fields coincide with the (non-canonically
normalized) mass eigenstates when we put them on mass-shell.  This
means that we can obtain the couplings of the KK states by simply
computing the holographic couplings on-shell, provided we also find
the correct normalization factors for the fields.

For the top field, the wave functions of the relevant on-shell (i.e.
$p=0$) components along the $\xi_u^+$ multiplet are given by
\be
\left\{
\begin{array}{l}
\xi_{u,L}^{+}(z) =\displaystyle -m_u z^{2-c_u} z_{\IR}^{c_u-c_{q_1}}
z_{\UV}^{c_{q_1}-5/2}\, t_L \equiv f_L^{+}(z)\, t_L\,,\\
\rule{0pt}{2.5em}\xi_{u,R}^{+}(z) =\displaystyle \frac{1}{\sqrt{z_{\UV}}}
\Big(\frac{z}{z_{\UV}}\Big)^{2+c_{u}}\, t_R \equiv f_R^{+}(z)\, t_R\,.
\end{array}
\right.
\label{xiuwavefunctions}
\ee
To find the correct normalization factors, $Z_{L,R}^+$, we must require
that the action have the canonical form for massless fields in a series
expansion in $p$ around the on-shell momentum.  The $L$ and $R$
normalization factors are thus given by
\be
(Z_{L,R}^+)^2 =\displaystyle \lim_{p\to 0} \frac{1}{z_{\UV}}\frac{\Pi_0^{q,u}}{p}
=\displaystyle \lim_{p\to 0} \frac{1}{z_{\UV}}\frac{\partial\Pi_0^{q,u}}
{\partial p}\,.
\ee

Due to the absence of zero modes, the holographic description of the
$\Z_2$-odd fields is more involved, since the action diverges for
$p\rightarrow 0$.  Nevertheless we can still use the holographic
approach by a suitable expansion around the on-shell momentum
$p=m_{odd}$, where $m_{odd}$ is a mass eigenvalue.

The NLOP states that couple to $t_R$ and $t_L$, $\psi_{-,R}$ and
$\psi_{-,L}$, are contained in the multiplet $\xi_u^-$ as in
Eq.~(\ref{fieldcomponents}), namely $\xi_R^- = (1,1)_R^{u-}$ and the
relevant component of $(2,2)_L^{u-}$, that we denote by $\xi_L^-$.
Both fields satisfy $(+-)$ b.c., so that we can choose as holographic
fields the $UV$ values of the corresponding components: $\psi_R^-
\equiv \sqrt{z_{\UV}}\,\xi_R^{-}(z_{\UV})$ and $\psi_L^- \equiv
\sqrt{z_{\UV}}\,\xi_L^{-}(z_{\UV})$.\footnote{In presence of EWSB,
when the various field components are mixed by the Higgs VEV, it is
better to choose holographic fields of the same chirality within a
multiplet (see \cite{Panico:2007qd}).  In the present case the two
components are independent of each other, so we can simply use
holographic fields with different chiralities.} The holographic
Lagrangian for this system at the quadratic level is
\be
{\cal L} = \overline\psi^{\,-}_R\frac{\pslash}{p}
\frac{\Pi_R}{z_{\UV}}\, \psi^-_R
+ \overline\psi^{\,-}_L\frac{\pslash}{p}\frac{\Pi_L}{z_{\UV}}\, \psi^-_L\,,
\ee
where $\Pi_L= -(\Pi_0+\Pi_1/2)^{-1}$ and $\Pi_R =\Pi_0$, evaluated
again with Minkowski signature.  The relevant on-shell field wave functions
are
\be
\left\{
\begin{array}{l}
\xi_R^{-}(z) =\displaystyle \frac{1}{\sqrt{z_{\UV}}}\Big(\frac{z}{z_{\UV}}\Big)^{5/2}
\frac{G_{--}(c_u, -i m_\Rs, z_{\IR}, z)}{G_{--}(c_u, -i m_\Rs, z_{\IR}, z_{\UV})}
\psi_R^- \equiv f_R^{-}(z)\, \psi_R^-\,,\\
\rule{0pt}{2em}\xi_L^{-} (z)=\displaystyle \frac{1}{\sqrt{z_{\UV}}}\Big(\frac{z}{z_{\UV}}\Big)^{5/2}
\frac{G_{++}(c_u, -i m_\Ls, z_{\IR}, z)}{G_{++}(c_u, -i m_\Ls, z_{\IR}, z_{\UV})}
\psi_L^- \equiv f_L^{-}(z)\, \psi_L^-\,.
\end{array}
\right.
\ee
The normalization factors $Z^{-}_{L,R}$ can be found by expanding the
action around the on-shell momentum.  One gets~\footnote{The
factor $1/2$ which appears in these expressions is due to the fact
that the fermions we are describing are massive.  In the holographic
approach the two chiralities of the fermions (say $\phi_{L,R}$) are
described by the same holographic field ($\widehat\phi_L$), so that
on-shell $\phi_L = \widehat\phi_L$ and $\phi_R = \slash{\!\!\!p}/p \,
\widehat\phi_L$.  In this case $\overline\phi(\, \pslash - m)\phi
\rightarrow \overline{\widehat\phi}_L 2 (p - m) \slash{\!\!\!p}/p\,
\widehat\phi_L$.}
\be
(Z^{-}_{L,R})^2 =\displaystyle \lim_{p\to m_{\Ls,\Rs}}
\frac{1}{2\, z_{\UV}}\frac{\Pi_{L, R}}{p - m_{\Ls,\Rs}}
= \lim_{p\to m_{\Ls,\Rs}}
\frac{1}{2\, z_{\UV}}\frac{\partial \Pi_{L, R}}{\partial p}\,,
\ee
where $m_{\Ls,\Rs}$ are the masses of the NLOP.

The holographic description of the $X_-$ gauge field is easily
obtained in an axial-type gauge $X_{-,5}=0$, as described in
\cite{Panico:2007qd}.  The quadratic holographic Lagrangian for the
transverse part of the $X_-$ gauge field is found to be
\be
{\cal L} = -\frac{1}{2\, z_{\UV}} \widehat X^t_{-,\mu} p\,\Pi_-\,
\widehat X_{-}^{t\,\mu}\,,
\ee
where $\widehat X^t_- \equiv \sqrt{z_{\UV}}\, X^t_-(z_{\UV})$ is the
holographic field.  The on-shell field wave function is
\be
X^t_{-,\mu} (z)= \displaystyle \frac{z}{z_{\UV}^{3/2}}
\frac{G_{++}(1/2, -im_{X_-}, z_{\IR}, z)}{G_{++}(1/2, -im_{X_-}, z_{\IR}, z_{\UV})}
\widehat X^t_- \equiv f_{X_-}(z)\, \widehat X^t_{-,\mu}\,,
\ee
and its normalization factor, $Z_{X_-}$, is
\be
Z_{X_-}^2 =\displaystyle \lim_{p\to m_{X_-}}
\frac{p}{z_{\UV}}\frac{\Pi_-}{p^2 - m_{X_-}^2}
= \lim_{p\to m_{X_-}}
\frac{\partial}{\partial p^2}\left(\frac{p\, \Pi_-}{z_{\UV}}\right)\,.
\ee

The interaction term can finally be written as
\bea
{\cal L}^{(3)} &=& \frac{2}{3}g_{5X} \int_{z_{\UV}}^{z_{\IR}} dz\, \Big(\frac{z_{\UV}}{z}\Big)^4
\left(\overline \xi_{L/R}^{\,-} \hspace{-2pt}\not \hspace{-2.7pt} X_- \xi_{L/R}^{+}
+ \rm{h.c.}\right)\nonumber\\
&=& \frac{2}{3}g_{5X}\, \overline \psi_{L/R} \not \hspace{-2.7pt} \widehat X_- t_{L/R}
\int _{z_{\UV}}^{z_{\IR}} dz \Big(\frac{z_{\UV}}{z}\Big)^4
\left(f_{L/R}^{-} f_{X_-} f_{L/R}^{+}\right)
 + \rm{h.c.}\,,
\eea
where $g_{5X}$ is the $5D$ $U(1)_X$ gauge coupling.  Taking into
account the normalization factors, we finally get:
\be
\tilde g_{L/R} = \frac{2}{3}\frac{g_{5X}}{Z_{L/R}^+ Z_{L/R}^{-} Z_{X_-}} \int_{z_{\UV}}^{z_{\IR}}
 dz\, \Big(\frac{z_{\UV}}{z}\Big)^4 \left(f_{L/R}^{-} f_{X-} f_{L/R}^{+}\right)\,.
 \label{gExplicit} 
\ee 
From the above relations it is possible to get an upper bound on
$\tilde g_L$ and $\tilde g_R$ by noticing that both couplings are
maximal when $t_R$ and $t_L$ reside mostly in the $\xi_u$ multiplet,
which occurs for $M_u\rightarrow \infty$ and $m_u\rightarrow \infty$,
respectively.\footnote{Notice that in the limit $M_u \rightarrow
\infty$, $m_u \rightarrow \infty$, the components of the $\xi_u$
multiplet related to $t_L$ and $t_R$ acquire $(++)$ b.c., while the
corresponding $\xi_{q_1}$ components satisfy $(+-)$ boundary
conditions [see also paragraph after Eq.~(\ref{app10})].} When
$M_u\rightarrow \infty$ and $c_u\simeq -1/2$, or $m_u\rightarrow
\infty$ and $c_u\simeq 1/2$ one gets:
\bea
 (Z_{R}^-)^2 & \simeq & (Z_{L}^-)^2  \simeq (Z_{X_-})^2\,, \hspace{2.4cm}
 f_{R}^-(z) \simeq  f_L^-(z) \simeq \left(\frac{z}{z_\UV}\right)^{3/2}\ f_{X_-}(z)\,, \nn \\
(Z_R^+)^2 & \simeq  & \log(z_\IR/z_\UV)\,, \hspace{3.2cm}
f_R^+(z) \simeq  \frac{1}{\sqrt{z_\UV}}\left(\frac{z}{z_\UV}\right)^{3/2}\,, \label{wavefunctionsZ} \\
(Z_L^+)^2&  \simeq & m_u^2 \left(\frac{z_\IR}{z_\UV}\right)^{1-2c_{q_1}}  \log(z_\IR/z_\UV)\, , \ \ \ \ 
f_L^+(z) \simeq - \frac{m_u}{\sqrt{z_\UV}}\left(\frac{z_\IR}{z_\UV}\right)^{1/2-c_{q_1}} \left(\frac{z}{z_\UV}\right)^{3/2}\,. \nn 
\eea
Using Eqs.~(\ref{wavefunctionsZ}), we see that the integral
over the wave functions in Eq.~(\ref{gExplicit}) reduces to the
normalization condition for the gauge field both for $\tilde g_R$ and
$\tilde g_L$, giving simply
\be
|\tilde g_{L/R}| \leq \frac{2}{3}\frac{g_{5X}}{\sqrt{z_\UV  \log(z_\IR/z_\UV)}} = \frac 23 g_X\,,
\ee
where in the latter equality we have used the relation between the 5D
coupling $g_{5X}$ and the 4D coupling $g_X$.  This proves
Eq.~(\ref{gxbound}) in the main text.

\end{document}